\documentclass[3p,10pt,a4paper,twoside,fleqn,sort&compress]{elsarticle}
\usepackage{amssymb,amsmath,latexsym}
\usepackage{pst-all}
\usepackage[varg]{pxfonts}
\usepackage{mathrsfs}
\newtheorem{theorem}{Theorem}[section]
\newtheorem{lemma}[theorem]{Lemma}

\newtheorem{example}[theorem]{Example}
\newtheorem{proposition}[theorem]{Proposition}

\def\endproof{\qed\endtrivlist}
\expandafter\let\csname endproof*\endcsname=\endproof

\def\qedsymbol{\ifmmode\bgroup\else$\bgroup\aftergroup$\fi
  \vcenter{\hrule\hbox{\vrule height.6em\kern.6em\vrule}\hrule}\egroup}
\def\qed{\ifmmode\else\unskip\nobreak\fi\quad\qedsymbol}

\renewcommand{\iff}{\Leftrightarrow}

\renewcommand{\le}{\leqslant}
\renewcommand{\ge}{\geqslant}

\newcommand{\im}{\qopname\relax{no}{Im}}

\begin{document}

\journal{\qquad}

\title{\Large\bf Algorithms for computing the greatest simulations and bisimulations\\ between fuzzy automata\tnoteref{t1}}
%%%
\tnotetext[t1]{Research supported by Ministry  of Science and Technological Development, Republic of Serbia, Grant No. 174013}
%%%
%%%%%%%%%%%%%%%%%%%%%%%%%%%%%%%%%%%%%%%%%%%%%%%%%%
\author[fsmun]{Miroslav \'Ciri\'c\corref{cor}}
\ead{miroslav.ciric@pmf.edu.rs}

\author[fsmun]{Jelena Ignjatovi\'c}
\ead{jelena.ignjatovic@pmf.edu.rs}

\author[fsmun]{Ivana Jan\v ci\'c}
\ead{ivanajancic84@gmail.com}

\author[tfc]{Nada Damljanovi\'c}
\ead{nada@tfc.kg.ac.rs}

\cortext[cor]{Corresponding author. Tel.: +38118224492; fax: +38118533014.}
\address[fsmun]{University of Ni\v s, Faculty of Sciences and Mathematics, Vi\v segradska 33, 18000 Ni\v s, Serbia}
\address[tfc]{University of Kragujevac, Technical faculty in \v Ca\v cak, Svetog Save 65, P. O. Box 131, 32000 \v Ca\v cak, Serbia}
%%%%%%%%%%%%%%%%%%%%%%%%%%%%%%%%%%%%%%%%%%%%%%%%%%

\begin{abstract}\small
Recently, two types of simulations (forward and backward simulations) and four types of bisimulations (forward,~back\-ward, forward-backward, and backward-forward bisimulations) between fuzzy automata have been introduced.~If~there is at least one~simu\-lation/bisimulation of some of these types between the given fuzzy automata,~it~has been proved that there is the greatest simulation/bisimulation of this kind.~In the present paper, for any of~the~above-mentioned types of simulations/bisimulations we provide an effective algorithm for deciding whether there is a simulation/bisimulation of this type between the given fuzzy automata, and for computing the greatest one, whenever it exists.~The algorithms are based on the~method~devel\-oped in [J. Ignjatovi\'c, M. \'Ciri\'c, S. Bogdanovi\'c, On the greatest solutions to certain systems of fuzzy relation inequalities and equations, Fuzzy Sets and Systems 161 (2010) 3081--3113], which comes down to the computing of the greatest post-fixed point, contained in a given fuzzy relation, of an isotone function on the lattice
of fuzzy~rela\-tions.

\end{abstract}


\maketitle

\section{Introduction}

Bisimulations have been introduced by Milner \cite{Milner.80} and Park \cite{Park.81} in computer science, where they~have been used to model equivalence between various~systems, as well as to reduce the number of states of~these systems.~Roughly at the same time they have been also discovered in some areas of mathematics, e.g., in modal logic and set theory.~They are employed today in a many areas of computer science,~such~as functional languages, object-oriented languages, types, data types, domains, databases, compiler optimi\-zations, program analysis, verification tools, etc.~For more information about bisimulations we refer to \cite{AILS.07,CL.08,DPP.04,GPP.03,LV.95,Milner.89,Milner.99,RM-C.00,Sang.09}.

The most common structures on which bisimulations have been studied are labelled transition systems, but they have also been investigated in the context of deterministic, nondeterministic, weighted,~probabil\-istic, timed and hybrid automata.~Recently, bisimulations have been discussed in the context of fuzzy~auto\-mata in \cite{CCK.11,CIDB.11,CSIP.07,CSIP.10,ICB.09,P.06,SCI.11,SLY.09}.~One can distinguish two general approaches to the concept of bisimulation for fuzzy automata.~The first approach, which we encounter in \cite{CCK.11,P.06,SLY.09}, uses ordinary crisp relations~and functions.~Another approach, proposed in \cite{CIDB.11,CSIP.07,CSIP.10,ICB.09,SCI.11}, is based on the use of fuzzy relations, which~have been shown to provide better results both in the state reduction and the modeling of equivalence of fuzzy automata.~In \cite{CSIP.07,CSIP.10} the state reduction of fuzzy automata was carried out by means of certain fuzzy~equi\-valences, which are exactly bisimulation fuzzy equivalences, and in \cite{SCI.11} it was proved that even~better results can be achieved by using suitable fuzzy quasi-orders, which are nothing but simulation fuzzy~quasi-orders.~Moreover, it turned out in \cite{CSIP.07,CSIP.10,SCI.11} that the state reduction problem for fuzzy automata is closely related to the problem of finding solutions to certain systems of fuzzy relation equations.~This enabled~not only to study fuzzy automata using very powerful tools of the theory of fuzzy sets, but also, it gave a~great contribution to the theory of fuzzy relational equations and has led to the development of the general theory of weakly linear fuzzy relation equations and inequalities in \cite{ICB.10}.

The same approach has been used in \cite{CIDB.11}, in the study of simulations and bisimulations between fuzzy automata, where simulations and bisimulations have also been defined as fuzzy relations.~There have been introduced two types of simulations (forward and backward simulations) and four types of bisimulations (forward, backward, forward-backward, and backward-forward bisimulations).~There has~been proved~that if there is at least one~simu\-lation/bisimulation of some of these types between the given fuzzy automata,~then there is the greatest simulation/bisimulation of this kind.~However, there has not been given~any effective algorithm for deciding whether there is a simulation/bisimulation of some of these types between the given fuzzy automata, and for computing the greatest one, if it exists.~In \cite{CIDB.11} a theorem has been proved which can be used just for checking the existence of a uniform forward bisimulation (i.e., a complete and surjective forward bisimulation) between the given fuzzy automata.~According to this theorem, there is a uniform forward bisimulation between fuzzy automata $\cal A$ and $\cal B$ if and only if there~is a special isomorphism between the~factor fuzzy automata of $\cal A$ and $\cal B$ with respect~to~their greatest forward bisimulation fuzzy equivalences.~Using the algorithm~provided in \cite{CSIP.10}, in numerous cases we can effectively compute the greatest forward bisimulation fuzzy equivalences $E$ on $\cal A$ and $F$ on $\cal B$.~Then we can construct factor fuzzy automata ${\cal A}/E$ and ${\cal B}/F$, and check whether there is an~isomorphism between them that satisfies an additional condition stated in \cite{CIDB.11}.~But, even when we~are able to effectively compute the greatest forward bisimulations $E$ and $F$ and construct the factor fuzzy automata ${\cal A}/E$ and ${\cal B}/F$, it may be difficult to determine whether there is an isomorphism between ${\cal A}/E$ and ${\cal B}/F$ that satisfies this additional condition.~This problem comes down to the well-known {\it graph isomorphism problem\/}, which is one of the~few important algorith\-mic problems whose rough computational com\-plexity~is still not known, and it is gener\-ally accepted that it lies between P and NP-complete if P$\ne $NP (cf.~\cite{Skiena.08}).~Fortunately, although no worst-case polynomial-time algorithm is known, testing graph isomorphism is usually not very hard in practice.

In this paper, for any of the above mentioned types of simulations/bisimulations we provide an effective algorithm for deciding whether there is a simulation/bisimulation of this type between the given fuzzy~auto\-mata, and for computing the greatest one, whenever it exists.~The algorithms are based on~the~method~devel\-oped in \cite{ICB.10}, which comes down to the computing of the greatest post-fixed point, contained in a given~fuzzy relation, of an isotone function on the lattice of fuzzy relations.~Namely, for each type of simulations and bisimulations we determine the corresponding isotone and image-localized function $\phi $ on the lattice of~fuzzy relations, as well as the corresponding initial relation $\psi $, and the computing of the greatest simulation/bisim\-ulation of this type we reduce to the the computing of the greatest post-fixed point of $\phi $ contained in~$\psi $. This is an iterative procedure by which we successively build a decreasing sequence of relations,~starting from the relation $\psi $ and using the function $\phi $.~If this sequence is finite, then it stabilizes and its smallest member is exactly the fuzzy relation which we are searching for, the greatest post-fixed point of $\phi $ contained in $\psi $.~We determine sufficient conditions under which this sequence is finite, when our algorithm terminates in the finite number of steps (cf.~Theorem \ref{th:alg}), as well as sufficient conditions under which the infimum of this sequence is exactly the fuzzy relation which we are searching for (cf.~Theorem \ref{th:inf}).~Modifying~the algorithms for computing the greatest simulations and bisimulations we provide algorithms for computing the greatest crisp simulations and bisimulations between fuzzy automata (cf.~Proposition \ref{prop:crisp.alg}).~These algorithms always terminate in a finite number of steps, independently of the properties
of the underlying structure of truth values, but we show that there are fuzzy automata such that there is a simulation or bisimulation of a given type between them, and there is not any crisp simulation/bisimulation of this type (cf.~Example \ref{ex:first}).

The structure of the paper is as follows.~In Section \ref{sec:prel} we give definitions of basic notions and notation~con\-cerning fuzzy sets and relations, and in Section \ref{sec:FFA} we define basic notions and notation concerning~fuzzy automata.~In Section \ref{sec:sim.bisim} we recall definitions of two types of simulations and four types of bisimulations introduced in \cite{CIDB.11}, and for any of them we define the corresponding function on the lattice of fuzzy relations. Section \ref{sec:main} contains our main results on the computing of the greatest simulations and bisimulations between fuzzy automata.~Examples presented in Section \ref{sec:ex} demonstrate the application of our algorithms and clarify relationships between different types of simulations and bisimulations.

It is worth noting that algorithms for deciding whether there are simulations and bisimulations between nondeterministic automata, and the computing of the greatest simulations and bisimulations, if they exist, were given in \cite{CIBJ.11,Kozen.97}.~Various algorithms for the computing of the greatest bisimulation equivalences on labelled transition systems can be found in \cite{DPP.04,GPP.03,KS.90,PT.87,RT.08,Saha.07}.

\section{Preliminaries}\label{sec:prel}

The terminology and basic notions in this section are according to \cite{Bel.02,BV.05,DP.80,DP.00,KY.95}.

We will use complete residuated lattices as the structures of membership (truth) values.~Residuated lattices are a
very general algebraic structure and generalize many  algebras with very important applications (see for example \cite{Bel.02,BV.05,Hajek.98,Hohle.95}).~A {\it residuated lattice\/} is an algebra ${\cal L}=(L,\land
,\lor , \otimes ,\to , 0, 1)$ such that
\begin{itemize}
\parskip=-2pt%\itemindent=3mm
\item[{\rm (L1)}] $(L,\land ,\lor , 0, 1)$ is a lattice with the least element $0$ and the
greatest element~$1$,
\item[{\rm (L2)}] $(L,\otimes ,1)$ is a commutative monoid with the unit $1$,
\item[{\rm (L3)}] $\otimes $ and $\to $ form an {\it adjoint pair\/}, i.e., they satisfy the
{\it adjunction property\/}: for all $x,y,z\in L$,
\begin{equation}\label{eq:adj}
x\otimes y \leqslant z \ \Leftrightarrow \ x\leqslant y\to z .
\end{equation}
\end{itemize}
If, in addition, $(L,\land ,\lor , 0, 1)$ is a complete lattice, then ${\cal L}$ is called a {\it
complete residuated lattice\/}.~Emphasizing their monoidal structure, in some sources residuated lattices are called
integral, commutative, residuated $\ell $-monoids \cite{Hohle.95}.

The operations $\otimes $ (called {\it multiplication\/}) and $\to
$ (called {\it residuum\/}) are intended for modeling the conjunction and implication of the
corresponding logical calculus, and supremum ($\bigvee $) and infimum ($\bigwedge $) are intended
for modeling of the existential and general quantifier, respectively. An operation $\leftrightarrow $ defined
by
\begin{equation}\label{eq:bires}
x\leftrightarrow y = (x\to y) \land (y\to x),
\end{equation}
called {\it biresiduum\/} (or {\it biimplication\/}), is used for modeling the equivalence of truth
values.

The most studied and applied structures of truth values, defined
on the real unit interval $[0,1]$ with $x\land y =\min
(x,y)$ and $x\lor y =\max (x,y)$, are the {\it {\L}ukasiewicz
structure\/} (where $x\otimes y = \max(x+y-1,0)$, $x\to y=
\min(1-x+y,1)$), the {\it Goguen} ({\it product\/}) {\it
structure\/} ($x\otimes y = x\cdot y$, $x\to y= 1$ if $x\le y$,
and~$=y/x$ otherwise), and the {\it G\"odel structure\/} ($x\otimes
y = \min(x,y)$, $x\to y= 1$ if $x\le y$, and $=y$
otherwise).~More~generally, an algebra $([0,1],\land ,\lor ,
\otimes,\to , 0, 1)$ is a complete~resi\-duated lattice if and
only if $\otimes $ is a left-continuous t-norm and the residuum is
defined by $x\to y = \bigvee \{u\in [0,1]\,|\, u\otimes x\le y\}$ (cf.~\cite{BV.05}).~Another
important set of truth values is the set
$\{a_0,a_1,\ldots,a_n\}$, $0=a_0<\dots <a_n=1$, with $a_k\otimes
a_l=a_{\max(k+l-n,0)}$ and $a_k\to a_l=a_{\min(n-k+l,n)}$. A
special case of the latter algebras is the two-element Boolean
algebra of classical logic with the support $\{0,1\}$.~The only
adjoint pair on the two-element Boolean algebra consists of the
classical conjunction and implication operations.~This structure
of truth values we call the {\it Boolean structure\/}.~A
residuated~lattice $\cal L$ satisfying $x\otimes y=x\land y$ is
called a {\it Heyting algebra\/}, whereas a Heyting algebra
satisfying the prelinearity axiom $(x\to y)\lor (y\to x)=1$ is
called a {\it G\"odel algebra\/}. If any finitelly
generated~sub\-algebra of a residuated lattice $\cal L$ is finite,
then $\cal L$ is called {\it locally finite\/}.~For example, every
G\"odel algebra, and hence, the G\"odel structure, is locally
finite, whereas the product structure is not locally finite.

If $\cal L$ is a complete residuated lattice, then for all $x,y,z\in L$ and any $\{y_i\}_{i\in
I}\subseteq L$ the following holds:
\begin{align}
& x\le y\ \ \text{implies}\ \ x\otimes z\le y\otimes z, \label{eq:is.1}\\
& x\le y\ \ \Leftrightarrow\ \ x\to y=1 ,\label{eq:x.le.y} \\
& x\otimes \bigvee_{i\in I}y_i = \bigvee_{i\in I}(x\otimes y_i),\label{eq:prod.sup} \\
& x\otimes \bigwedge_{i\in I}y_i \le  \bigwedge_{i\in I}(x\otimes y_i), \label{eq:prod.inf} \end{align}
For other properties of complete residuated lattices we refer to \cite{Bel.02,BV.05}.

In the further text $\cal L$ will be a complete residuated lattice.~A {\it
fuzzy subset\/} of a set $A$ {\it over\/} ${\cal L}$, or~simply a {\it fuzzy
subset\/} of $A$, is any function from $A$ into $L$.~Ordinary crisp subsets~of~$A$ are considered as fuzzy subsets of $A$ taking membership values in the set
$\{0,1\}\subseteq L$.~Let $f$ and $g$ be two
fuzzy subsets of $A$.~The {\it equality\/} of $f$ and $g$ is defined as the
usual equality of functions, i.e., $f=g$ if and only if $f(x)=g(x)$, for every
$x\in A$. The {\it inclusion\/} $f\leqslant g$ is also defined pointwise:~$f\leqslant g$ if
and only if $f(x)\leqslant g(x)$, for every $x\in A$.~Endowed with this partial order the set ${\cal F}(A)$ of all fuzzy subsets of $A$ forms a complete residuated lattice, in which the
meet (intersection) $\bigwedge_{i\in I}f_i$ and the join (union) $\bigvee_{i\in I}f_i$ of an
arbitrary family $\{f_i\}_{i\in I}$ of fuzzy subsets of $A$ are functions from $A$ into $L$
defined by
\[
\left(\bigwedge_{i\in I}f_i\right)(x)=\bigwedge_{i\in I}f_i(x), \qquad \left(\bigvee_{i\in
I}f_i\right)(x)=\bigvee_{i\in I}f_i(x),
\]
and the {\it product\/} $f\otimes g$ is a fuzzy subset defined by $f\otimes g
(x)=f(x)\otimes g(x)$, for every $x\in A$.~

Let $A$ and $B$ be non-empty sets.~A {\it fuzzy relation between sets\/} $A$ and $B$ (or a {\it fuzzy relation from $A$ to $B$\/}) is any function from $A\times B$ into~$L$, that is to say, any fuzzy subset of $A\times B$, and the equality, inclusion (ordering), joins and meets of fuzzy relations
are defined as for fuzzy sets.~In particular, a {\it fuzzy relation on a set\/} $A$ is any function from $A\times A$ into $L$, i.e., any fuzzy subset of $A\times A$.~To highlight the difference between fuzzy relations between two sets and those on a set, fuzzy relations between two sets will be usually denoted by small Greek letters, and fuzzy relations on a set by capital Latin letters.~The set of all fuzzy relations from~$A$ to $B$ will be denoted by
${\cal R}(A,B)$, and the set of all fuzzy relations on a set $A$ will be
denoted by ${\cal R}(A)$.~The {\it converse\/} (in some sources called
{\it inverse\/} or {\it transpose\/}) of a fuzzy relation $\varphi \in {\cal R}(A,B)$ is a fuzzy relation
$\varphi^{-1}\in {\cal R}(B,A)$ defined by $\varphi^{-1}(b,a)=\varphi (a,b)$, for all $a\in A$ and $b\in B$.~A {\it crisp relation\/} is a fuzzy relation which takes values only in the set $\{0,1\}$, and if $\varphi $ is a crisp relation of $A$ to $B$, then expressions ``$\varphi (a,b)=1$'' and ``$(a,b)\in \varphi $'' will have the same meaning.

For non-empty sets $A$, $B$ and $C$, and fuzzy relations $\varphi\in {\cal R}(A, B)$ and $\psi \in {\cal R}(B, C)$, their {\it composition\/}~$\varphi\circ \psi$ is an fuzzy relation from ${\cal R}(A, C)$ defined by
\begin{equation}\label{eq:comp.rr}
(\varphi \circ \psi )(a,c)=\bigvee_{b\in B}\,\varphi(a,b)\otimes \psi(b,c),
\end{equation}
for all $a\in A$ and $c\in C$.~If $\varphi $ and $\psi $ are crisp relations, then $\varphi\circ \psi $ is an ordinary composition of relations,~i.e.,
\[
\varphi\circ \psi=\left\{(a,c)\in A\times C\mid (\exists b\in B)\, (a,b)\in \varphi\ \&\ (b,c)\in \psi\right\},
\]
and if $\varphi $ and $\psi $ are functions, then $\varphi\circ \psi $ is an ordinary composition of functions, i.e.,
$(\varphi\circ \psi )(a)=\psi (\varphi(a))$, for every $a\in A$.~Next, if $f\in {\cal F}(A)$, $\varphi\in {\cal R}(A, B)$ and $g\in {\cal F}(B)$, the {\it compositions\/} $f\circ \varphi$ and $\varphi\circ g$ are fuzzy subsets of $B$ and $A$, respectively, which are defined by
\begin{equation}\label{eq:comp.sr}
(f \circ \varphi)(b)=\bigvee_{a\in A}\,f(a)\otimes \varphi(a,b),\quad
(\varphi \circ g)(a)=\bigvee_{b\in B}\,\varphi(a,b)\otimes g(b),
\end{equation}
for every $a\in A$ and $b\in B$.

In particular, for $f,g\in {\cal F}(A)$ we write
\begin{equation}\label{eq:comp.ss}
f \circ g =\bigvee_{a\in A}\,f(a)\otimes g(a) .
\end{equation}
The value $f\circ g$ can be interpreted as the "degree of overlapping" of $f$ and $g$.~In particular, if $f$ and $g$ are crisp sets and $\varphi $ is a crisp relation, then
\[
f \circ \varphi = \left\{b\in B\mid (\exists a\in f)\, (a,b)\in\varphi \right\}, \ \ \
\varphi\circ g = \left\{a\in A\mid (\exists b\in g)\, (a,b)\in\varphi \right\}.
\]

Let $A$, $B$, $C$ and $D$ be non-empty sets. Then for any $\varphi_1\in {\cal R}(A, B)$, $\varphi_2\in {\cal R}(B,C)$ and $\varphi_3\in {\cal R}(C, D)$~we~have
\begin{align}
&(\varphi_{1}\circ \varphi_{2})\circ \varphi_{3} = \varphi_{1}\circ (\varphi_{2}\circ \varphi_{3}), \label{eq:comp.as}
\end{align}
and for $\varphi_{0}\in {\cal R}(A, B)$,\, $\varphi_{1},\varphi_{2}\in {\cal R}(B,C)$ and $\varphi_{3}\in {\cal R}(C, D)$~we~have that
\begin{align}
&\varphi_{1}\leqslant \varphi_{2}\ \ \text{implies}\ \ \varphi_1^{-1}\leqslant \varphi_2^{-1},\ \ \varphi_{0}\circ \varphi_{1} \leqslant  \varphi_{0}\circ \varphi_{2}, \ \ \text{and}\ \ \varphi_{1}\circ \varphi_{3} \leqslant  \varphi_{2}\circ \varphi_{3}. \label{eq:comp.mon}
\end{align}
Further, for any $\varphi\in {\cal R}(A, B)$, $\psi\in {\cal R}(B, C)$, $f\in {\cal F}(A)$, $g\in {\cal F}(B)$ and $h\in {\cal F}(C)$
we can easily verify~that
\begin{equation}\label{eq:rs.comp.as}
(f\circ \varphi)\circ \psi=f\circ (\varphi\circ \psi), \quad (f\circ \varphi)\circ g=f\circ (\varphi\circ g),\quad (\varphi\circ \psi)\circ h=\varphi\circ (\psi\circ h)
\end{equation}
and consequently, the parentheses~in (\ref{eq:rs.comp.as}) can be omitted, as well as the parentheses~in (\ref{eq:comp.as}).

Finally, for all $\varphi,\varphi_i\in {\cal R}(A, B)$ ($i\in I$) and $\psi,\psi_i\in {\cal R}(B, C)$ ($i\in I$) we have that
\begin{align}
&(\varphi\circ \psi)^{-1} = \psi^{-1}\circ \varphi^{-1}, \label{eq:comp.inv} \\
& \varphi\circ \bigl(\bigvee_{i\in I}\psi_i\bigr) = \bigvee_{i\in I}(\varphi\circ \psi_i) , \ \
\bigl(\bigvee_{i\in I}\varphi_i\bigr)\circ \psi = \bigvee_{i\in I}(\varphi_i\circ \psi), \label{eq:comp.sup} \\
%%%%%
%%%& \varphi\circ \bigl(\bigwedge_{i\in I}\psi_i\bigr) \leqslant \bigwedge_{i\in I}(\varphi\circ \psi_i) , \ \
%%%\bigl(\bigwedge_{i\in I}\varphi_i\bigr)\circ \psi \leqslant \bigwedge_{i\in I}(\varphi_i\circ \psi). \label{eq:comp.inf} \\
&\bigl(\bigvee_{i\in I}\varphi_i\bigr)^{-1} = \bigvee_{i\in I}\varphi_i^{-1}. \label{eq:sup.inv}
\end{align}

We note that if $A$, $B$ and $C$ are finite sets of cardinality $|A|=k$, $|B|=m$ and $|C|=n$, then $\varphi \in {\cal R}(A, B)$~and $\psi \in {\cal R}(B, C)$ can be treated as $k\times m$ and $m\times n$ fuzzy matrices over $\cal L$, and $\varphi\circ \psi$ is the matrix~pro\-duct.~Analo\-gously, for $f\in {\cal F}(A)$ and $g\in {\cal F}(B)$ we can treat $f\circ \varphi$ as the product of a $1\times k$ matrix $f$ and a $k\times m$~matrix $\varphi $ (vector-matrix product),
$\varphi\circ g$ as the product of an $k\times m$ matrix $\varphi$ and an $m\times 1$ matrix $g^t$,
the transpose~of~$g$ (matrix-vector product), and $f\circ g$ as the scalar product of vectors $f$ and $g$.

.

\section{Fuzzy automata}\label{sec:FFA}

In this paper we study fuzzy automata with membership values in complete residuated lattices.

In the rest of the paper, if not noted otherwise, let ${\cal L}$ be a complete residuated lattice and let $X$ be an (finite) alphabet.
A {\it fuzzy automaton over\/} $\cal L$ and $X$ , or simply a {\it fuzzy automaton\/}, is a quadruple
${\cal A}=(A,\delta^A,\sigma^A,\tau^A )$, where~$A$ is a non-empty set, called  the
{\it set of states\/}, $\delta^A:A\times X\times A\to L$~is~a
fuzzy subset~of $A\times X\times A$, called the {\it fuzzy transition function\/}, and $\sigma^A: A\to L$ and $\tau^A : A\to L$ are the fuzzy subsets of~$A$, called the {\it fuzzy set of initial states} and the {\it fuzzy set of terminal states}, respectively.~We can
interpret $\delta^A (a,x,b)$ as the degree~to~which an~input letter $x\in X$~causes~a~transition from a state $a\in A$ into a
state $b\in A$, whereas we can interpret $\sigma^A(a)$ and $\tau^A(a)$ as the degrees to which $a$ is respectively an input state and a terminal state.

Many results presented in this paper can also be proved for fuzzy automata with infinite sets of states. However, here we deal with effective algorithms for computing the greatest simulations and bisimulations, and for that reason we will assume that the sets of states of the considered fuzzy automata are always \emph{finite\/}.

Let $X^*$ denote the free monoid over the alphabet $X$, and  let $\varepsilon\in X^*$ be the
empty word. The function~$\delta^A $ can be extended~up to a function
$\delta^A_*:A\times X^*\times A\to L$ as follows: If $a,b\in A$, then
\begin{equation}\label{eq:delta.e}
\delta^A_*(a,\varepsilon ,b)=\begin{cases}\ 1, & \text{if}\ a=b, \\ \ 0, & \mbox{otherwise,}
\end{cases}
\end{equation}
and if $a,b\in A$, $u\in X^*$ and $x\in X$, then
\begin{equation}\label{eq:delta.x}
\delta^A_*(a,ux ,b)= \bigvee _{c\in A} \delta^A_*(a,u,c)\otimes \delta^A (c,x,b).
\end{equation}

By (\ref{eq:prod.sup}) and Theorem 3.1 \cite{LP.05} (see also \cite{Qiu.01,Qiu.02,Qiu.06}), we have that
\begin{equation}\label{eq:delta.uv}
\delta^A_*(a,uv,b)= \bigvee _{c\in A} \delta^A_*(a,u,c)\otimes \delta^A_*(c,v,b),
\end{equation}
for all $a,b\in A$ and $u,v\in X^*$, i.e., if $w=x_1\cdots x_n$, for $x_1,\ldots ,x_n\in X$,
then
\begin{equation}\label{eq:delta.x1xn}
\delta^A_*(a,w,b)= \bigvee _{(c_1,\ldots ,c_{n-1})\in A^{n-1}}
\delta^A(a,x_1,c_1)\otimes \delta^A(c_1,x_2,c_2) \otimes\cdots \otimes \delta^A(c_{n-1},x_n,b).
\end{equation}
Intuitively, the product $\delta^A(a,x_1,c_1)\otimes \delta^A(c_1,x_2,c_2) \otimes\cdots \otimes
\delta^A(c_{n-1},x_n,b)$ represents the degree to which the input word $w$ causes a transition from a
state $a$ into a state $b$ through the sequence of intermediate states $c_1,\ldots ,c_{n-1}\in A$,
and $\delta^A_*(a,w,b)$ represents the supremum of degrees of all possible transitions from $a$ into
$b$ caused by $w$. Also, we can visualize~a fuzzy finite automaton $\cal A$ representing it as a
labelled directed graph whose nodes are~states of $\cal A$, and an edge from a node $a$ into a node
$b$ is labelled by pairs of the form $x/\delta^{A}(a,x,b)$, for any $x\in X$.

It is worth noting that fuzzy automata are a natural generalization of nondeterministic
and deterministic automata.~Namely, if $\delta^A $ is a crisp subset of $A\times X\times A$, that is, $\delta^A:A\times X\times A\to
\{0,1\}$, and $\sigma^A$ and $\tau^A$ are crisp subsets~of~$A$, then $\cal A$ is an ordinary
{\it nondeterministic automaton\/}.~In other words, nondeterministic automata are fuzzy automata
over the Boolean structure.~They will also be called {\it Boolean automata\/}.~If $\delta^A $ is a function of $A\times X$ into~$A$, $\sigma^A$ is
a one-element crisp subset of $A$, that is, $\sigma^A=\{a_0\}$, for some $a_0\in A$, and $\tau^A$ is a fuzzy subset of $A$, then
$\cal A$ is called a {\it deterministic fuzzy automaton\/}, and it is denoted by ${\cal A}=(A,\delta^A,a_0,\tau^A )$.~In \cite{CDIV.10,DSV.10} the name {\it crisp-deterministic\/} was used.~For more information on deterministic fuzzy automata we refer to \cite{Bel.02a,IC.10,ICB.08,ICBP.10,JIC.11,LP.05}.~Evidently, if $\delta^A$ is a crisp subset of $A\times X\times A$, or a function of $A\times X$ into $A$, then $\delta^A_*$ is also a crisp subset of $A\times X^*\times A$, or a function of $A\times X^*$ into $A$, respectively.~A~deterministic fuzzy automaton ${\cal A}=(A,\delta^A,a_0,\tau^A )$, where $\tau^A$ is a crisp subset of $A$, is an ordinary deterministic automaton.

If for any $u\in X^*$ we define a fuzzy relation $\delta^A_u$ on $A$ by
\begin{equation}\label{eq:trans.rel}
\delta^A_u (a,b) = \delta^A_* (a,u,b) ,
\end{equation}
for all $a,b\in A$, called the {\it fuzzy transition relation\/} determined by $u$, then (\ref{eq:delta.uv}) can be written as
\begin{equation}\label{eq:delta.uv.2}
\delta^A_{uv}= \delta^A_u\circ \delta^A_v,
\end{equation}
for all $u,v\in X^*$.

The {\it reverse fuzzy automaton} of a fuzzy automaton ${\cal A}=(A,\delta^A, \sigma^A ,\tau^A )$ is defined as the fuzzy automaton $\bar{{\cal A}}=(A,\bar{\delta}^A,\bar{\sigma}^A
,\bar{\tau}^A )$ whose fuzzy transition function and fuzzy sets of initial and terminal states are defined by
$\bar{\delta}^A(a_1,x,a_2)=\delta^A(a_2,x,a_1)$ for all $a_1,a_2\in A$ and $x\in X$, $\bar{\sigma}^A=\tau^A$ and $\bar{\tau}^A=\sigma^A$.~In other words, $\bar{\delta}_x^A= (\delta_x^A)^{-1}$, for each $x\in X$.

A {\it fuzzy language\/} in $X^*$ over ${\cal L}$, or
briefly a {\it fuzzy language\/}, is any fuzzy subset of~$X^*$, i.e., any function from
$X^*$ into $L$.~A {\it fuzzy language recognized by a fuzzy automaton\/} ${\cal A}=(A,\delta^A , \sigma^A,\tau^A )$,
denoted as $L({\cal A})$, is a fuzzy language in ${\cal F}(X^*)$ defined by
\begin{equation}\label{eq:recog}
L({\cal A})(u) = \bigvee_{a,b\in A} \sigma^{A} (a)\otimes \delta_*^A(a,u,b)\otimes \tau^A(b) ,
\end{equation}
or equivalently,
\begin{equation}\label{eq:recog.comp}
\begin{aligned}
L({\cal A})(e) &= \sigma^A\circ \tau^A ,\\
L({\cal A})(u) &= \sigma^A\circ \delta_{x_1}^A\circ \delta_{x_2}^A\circ \cdots
\circ \delta_{x_n}^A\circ \tau^A ,
\end{aligned}
\end{equation}
for any $u=x_1x_2\dots x_n\in X^+$, where $x_1,x_2,\ldots ,x_n\in X$.~In other words, the equality (\ref{eq:recog}) means that the~membership degree of the word
$u$~to~the fuzzy language $L({\cal A})$ is equal to the degree to which $\cal A$ recognizes or accepts
the word $u$.~Using notation from (\ref{eq:comp.sr}), and
the second equality in (\ref{eq:rs.comp.as}), we can state (\ref{eq:recog}) as
\begin{equation}\label{eq:recog.comp.2}
L({\cal A})(u) = \sigma^A \circ \delta^A_u\circ \tau^A .
\end{equation}
Fuzzy automata $\cal A$~and~$\cal B$ are called {\it language-equivalent\/}, or sometimes just {\it equivalent\/}, if $L({\cal A})=L({\cal B})$.

For more information on fuzzy automata with membership values in complete residuated lattices we refer to the recent papers \cite{CSIP.07,CSIP.10,ICB.08,ICB.09,ICBP.10,Qiu.01,Qiu.02,Qiu.06,SCI.11,WQ.10,XQ.09a,XQ.09b,XQL.09,XQLF.07}, and for information on fuzzy automata over related types of lattices we refer to \cite{IC.10,LP.05,P.04a,P.04b,PK.04,PZ.08}.~The most complete overview of the results concerning the ``classical'' fuzzy automata taking membership values in the G\"odel or the Goguen structure can be found in the book \cite{MM.02}.

\section{Simulations and bisimulations between fuzzy automata}\label{sec:sim.bisim}

First we recall definitions of two types of simulations and four types of bisimulations introduced in \cite{CIDB.11}.

Let ${\cal A}=(A,\delta^A,\sigma^A,\tau^A)$ and ${\cal B}=(B,\delta^B,\sigma^B,\tau^B)$ be fuzzy automata, and let $\varphi\in {\cal R}(A,B)$ be a non-empty fuzzy relation. We call $\varphi $ a {\it forward simulation\/} if it satisfies
%%%%%%%%%%%%%%%%%%%%%%%%%%%%%%%%%%%%%%%%%%%%%%%%%%%%%%%%%%%%%%%%%%
\begin{align}\label{eq:fs-s}
&\sigma^A\leqslant \sigma^B\circ\varphi^{-1}, \tag{$fs$-1}\\
\label{eq:fs-d}
&\varphi^{-1}\circ \delta_x^A\leqslant \delta_x^B\circ \varphi^{-1},\quad \text{for every $x\in X$}, \tag{$fs$-2} \\
\label{eq:fs-t}
&\varphi^{-1}\circ\tau^A\leqslant \tau^B,\tag{$fs$-3}
\end{align}
%%%%%%%%%%%%%%%%%%%%%%%%%%%%%%%%%%%%%%%%%%%%%%%%%%%%%%%%%%%%%%%%%%
and a {\it backward simulation\/} if
%%%%%%%%%%%%%%%%%%%%%%%%%%%%%%%%%%%%%%%%%%%%%%%%%%%%%%%%%%%%%%%%%%
\begin{align}\label{eq:bs-s}
&\tau^A\leqslant\varphi\circ\tau^B,\tag{$bs$-1}\\
\label{eq:bs-d}
&\delta_x^A\circ \varphi \leqslant \varphi\circ \delta_x^B ,\quad \text{for every $x\in X$},\tag{$bs$-2}\\
\label{eq:bs-t}
&\sigma^A\circ\varphi\leqslant \sigma^B.\tag{$bs$-3}
\end{align}
%%%%%%%%%%%%%%%%%%%%%%%%%%%%%%%%%%%%%%%%%%%%%%%%%%%%%%%%%%%%%%%%%%
Furthermore, we call $\varphi $ a {\it forward bisimulation\/} if both $\varphi $ and $\varphi^{-1}$ are forward simulations, i.e., if $\varphi $ satisfies
%%%%%%%%%%%%%%%%%%%%%%%%%%%%%%%%%%%%%%%%%%%%%%%%%%%%%%%%%%%%%%%%%%
\begin{align}\label{eq:fb-s}
&\sigma^A\leqslant \sigma^B\circ\varphi^{-1}, &&\sigma^B\leqslant \sigma^A\circ\varphi, &\hspace{50mm}&\tag{$fb$-1}\\
\label{eq:fb-d}
&\varphi^{-1}\circ \delta_x^A\leqslant \delta_x^B\circ \varphi^{-1},&&\varphi\circ \delta_x^B\leqslant \delta_x^A\circ \varphi,\quad \text{for every $x\in X$}, &&\tag{$fb$-2}\\
\label{eq:fb-t}
&\varphi^{-1}\circ\tau^A\leqslant \tau^B, &&\varphi\circ\tau^B\leqslant \tau^A, &&\tag{$fb$-3}
\end{align}
%%%%%%%%%%%%%%%%%%%%%%%%%%%%%%%%%%%%%%%%%%%%%%%%%%%%%%%%%%%%%%%%%%
and a {\it backward bisimulation\/}, if both $\varphi $ and $\varphi^{-1}$ are backward simulations, i.e., if
$\varphi $ satisfies
%%%%%%%%%%%%%%%%%%%%%%%%%%%%%%%%%%%%%%%%%%%%%%%%%%%%%%%%%%%%%%%%%%
\begin{align}\label{eq:bb-s}
&\tau^A\leqslant\varphi\circ\tau^B,&&\tau^B\leqslant\varphi^{-1}\circ\tau^A, &\hspace{50mm}&\tag{$bb$-1}\\
\label{eq:bb-d}
&\delta_x^A\circ \varphi \leqslant \varphi\circ \delta_x^B ,&&\delta_x^B\circ \varphi^{-1} \leqslant \varphi^{-1}\circ \delta_x^A ,\quad \text{for every $x\in X$}, &&\tag{$bb$-2}\\
\label{eq:bb-t}
&\sigma^A\circ\varphi\leqslant \sigma^B,&&\sigma^B\circ\varphi^{-1}\leqslant \sigma^A. &&\tag{$bb$-3}
\end{align}
%%%%%%%%%%%%%%%%%%%%%%%%%%%%%%%%%%%%%%%%%%%%%%%%%%%%%%%%%%%%%%%%%%
Also, if $\varphi $ is a forward simulation and $\varphi^{-1}$ is a backward simulation, i.e., if $\varphi $ satisfies
%%%%%%%%%%%%%%%%%%%%%%%%%%%%%%%%%%%%%%%%%%%%%%%%%%%%%%%%%%%%%%%%%%
\begin{align}\label{eq:fbb-s}
&\sigma^A\le \sigma^B\circ\varphi^{-1}, &\tau^B\le \varphi^{-1}\circ \tau^A
,  &&\tag{$fbb$-1} \\
\label{eq:fbb-d}
&\varphi^{-1}\circ \delta_x^A= \delta_x^B\circ \varphi^{-1}, & && \text{for every $x\in X$},\hspace{50mm} \tag{$fbb$-2}\\
\label{eq:fbb-t}
&\sigma^B\circ\varphi^{-1}\le \sigma^A, &\varphi^{-1}\circ\tau^A\le  \tau^B,
&& \tag{$fbb$-3}
\end{align}
%%%%%%%%%%%%%%%%%%%%%%%%%%%%%%%%%%%%%%%%%%%%%%%%%%%%%%%%%%%%%%%%%%
then $\varphi $ is called a {\it forward-backward bisimulation\/}, and if $\varphi $ is a backward simulation and $\varphi^{-1}$ is a forward simulation, i.e., if
%%%%%%%%%%%%%%%%%%%%%%%%%%%%%%%%%%%%%%%%%%%%%%%%%%%%%%%%%%%%%%%%%%
\begin{align}\label{eq:bfb-s}
&\sigma^B\le \sigma^A\circ\varphi ,& \tau^A\le \varphi\circ\tau^B , && \tag{$bfb$-1} \\
\label{eq:bfb-d}
&\delta_x^A\circ \varphi = \varphi\circ \delta_x^B ,&& &\text{for every $x\in X$}, \hspace{50mm} \tag{$bfb$-2} \\
\label{eq:bfb-t}
&\sigma^A\circ\varphi \le \sigma^B &\varphi\circ\tau^B\le\tau^A. &&\tag{$bfb$-3}
\end{align}
%%%%%%%%%%%%%%%%%%%%%%%%%%%%%%%%%%%%%%%%%%%%%%%%%%%%%%%%%%%%%%%%%%
then $\varphi $ is called a {\it backward-forward bisimulation\/}.

For the sake of simplicity, we~will call $\varphi $ just a {\it simulation\/} if $\varphi $ is either a forward or a backward simulation, and just a {\it bisimulation\/} if $\varphi $ is any of the four types of bisimulations defined above.~Moreover, forward and backward bisimulations will be called {\it homotypic\/}, whereas backward-forward and forward-backward bisimulations will be called {\it heterotypic\/}.

%%%
\begin{figure}
%%%%
\begin{center}
%%%%
\psset{unit=0.4cm}
%\newpsobject{showgrid}{psgrid}{subgriddiv=1,griddots=10,gridlabels=6pt}\showgrid
%%%
\begin{pspicture}(-10,-7.2)(10,5)%\showgrid
%%%%
\psarc[linecolor=lightgray,linewidth=1pt](-5,6){2.4}{215}{325}
\psarc[linecolor=lightgray,linewidth=1pt](-5,-8){2.4}{35}{145}
\psarc[linecolor=lightgray,linewidth=1pt](5,6){2.4}{215}{325}
\psarc[linecolor=lightgray,linewidth=1pt](5,-8){2.4}{35}{145}
%%%%
\psellipse[linecolor=gray,linewidth=1.5pt](-5,-1)(4,6.5)
\psellipse[linecolor=gray,linewidth=1.5pt](5,-1)(4,6.5)
%%%%
\rput(-9.1,2.7){\Large $A$}
\rput(9.1,2.7){\Large $B$}
\rput(-7.2,5.3){\large $\sigma^A$}
\rput(7.5,5.3){\large $\sigma^B$}
\rput(-7.4,-7.2){\large $\tau^A$}
\rput(7.0,-7.2){\large $\tau^B$}
%%%%
\pnode(-5,6){AI}
\cnode(-5,4.6){.55}{A0}
\rput(A0){\scriptsize $a_0$}
\cnode[linewidth=0,linecolor=white](-5,2.5){.55}{A1}
\rput(-5,2.7){$\vdots$}
\cnode(-5,0.6){.55}{Ak}
\rput(Ak){\scriptsize $a_k$}
\cnode(-5,-2.5){.6}{Ak1}
\rput(Ak1){\scriptsize $a_{k+1}$}
\cnode[linewidth=0,linecolor=white](-5,-4.5){.55}{Ak2}
\rput(-5,-4.3){$\vdots$}
\cnode[doubleline=true,doublesep=1.5pt](-5,-6.6){.55}{An}
\rput(An){\scriptsize $a_n$}
%%%%
\pnode(5,6){BI}
\cnode[linestyle=dashed,dash=3pt 2pt](5,4.6){.55}{B0}
\rput(B0){\scriptsize $b_0$}
\cnode[linewidth=0,linecolor=white](5,2.5){.55}{B1}
\rput(5,2.7){$\vdots$}
\cnode[linestyle=dashed,dash=3pt 2pt](5,0.5){.55}{Bk}
\rput(Bk){\scriptsize $b_k$}
\cnode[linestyle=dashed,dash=3pt 2pt](5,-2.5){.6}{Bk1}
\rput(Bk1){\scriptsize $b_{k+1}$}
\cnode[linewidth=0,linecolor=white](5,-4.5){.55}{Bk2}
\rput(5,-4.3){$\vdots$}
\cnode[doubleline=true,doublesep=1.5pt,linestyle=dashed,dash=3pt 2pt](5,-6.6){.55}{Bn}
\rput(Bn){\scriptsize $b_n$}
%%%%
\ncline{->}{AI}{A0}
\ncarc[arcangle=-30]{->}{A0}{A1}\Bput[2pt]{\scriptsize $x_1$}
\ncarc[arcangle=-30]{->}{A1}{Ak}\Bput[2pt]{\scriptsize $x_k$}
\ncarc[arcangle=-30]{->}{Ak}{Ak1}\Bput[2pt]{\scriptsize $x_{k+1}$}
\ncarc[arcangle=-30]{->}{Ak1}{Ak2}\Bput[2pt]{\scriptsize $x_{k+2}$}
\ncarc[arcangle=-30]{->}{Ak2}{An}\Bput[2pt]{\scriptsize $x_n$}
%%%%
\ncline[linestyle=dashed,dash=3pt 2pt]{->}{BI}{B0}
\ncarc[arcangle=30,linestyle=dashed,dash=3pt 2pt]{->}{B0}{B1}\Aput[2pt]{\scriptsize $x_1$}
\ncarc[arcangle=30,linestyle=dashed,dash=3pt 2pt]{->}{B1}{Bk}\Aput[2pt]{\scriptsize $x_k$}
\ncarc[arcangle=30,linestyle=dashed,dash=3pt 2pt]{->}{Bk}{Bk1}\Aput[2pt]{\scriptsize $x_{k+1}$}
\ncarc[arcangle=30,linestyle=dashed,dash=3pt 2pt]{->}{Bk1}{Bk2}\Aput[2pt]{\scriptsize $x_{k+2}$}
\ncarc[arcangle=30,linestyle=dashed,dash=3pt 2pt]{->}{Bk2}{Bn}\Aput[2pt]{\scriptsize $x_n$}
%%%%
\ncarc[arcangle=30,linestyle=dashed,dash=3pt 2pt]{->}{A0}{B0}\Aput[3pt]{$\varphi$}
\ncarc[arcangle=30,linestyle=dashed,dash=3pt 2pt]{->}{Ak}{Bk}
\ncarc[arcangle=30,linestyle=dashed,dash=3pt 2pt]{->}{Ak1}{Bk1}
\ncarc[arcangle=30,linestyle=dashed,dash=3pt 2pt]{->}{An}{Bn}
\end{pspicture}\\
\caption{Forward and backward simulation}\label{fig:FBS}
%%%%
\end{center}
%%%%
\end{figure}
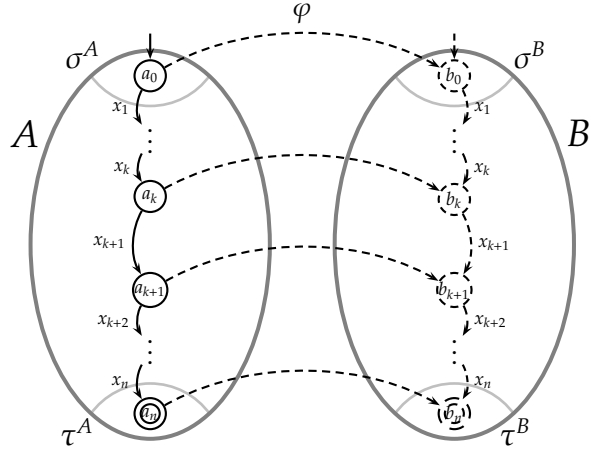
%%%%

The meaning of forward and backward simulations can be best explained in the case when $\cal A$ and~$\cal B$~are  nondeterministic (Boolean) automata.~For this purpose we will use the diagram shown in Figure 1.~Let~$\varphi $~be~a forward simulation between $\cal A$ and $\cal B$ and let $a_0,a_1,\ldots ,a_n$ be an arbitrary successful run of the automaton $\cal A$ on a word $u=x_1x_2\cdots x_n$ ($x_1,x_2,\ldots ,x_n\in X$), i.e., a sequence of states of $\cal A$ such that $a_0\in \sigma^A$,
$(a_k,a_{k+1})\in \delta^A_{x_{k+1}}$, for $0\leqslant k\leqslant n-1$, and $a_n\in \tau^A$.~According to (\ref{eq:fs-s}), there is an initial state $b_0\in \sigma^B$ such that $(a_0,b_0)\in\varphi $.~Suppose that for some $k$, $0\le k\le n-1$, we have built a sequence of states $b_0,b_1,\ldots ,b_k$ such that
$(b_{i-1},b_i)\in \delta^B_{x_i}$ and $(a_i,b_i)\in \varphi $, for each $i$, $1\le i\le k$.~Then $(b_k,a_{k+1})\in \varphi^{-1}\circ \delta^A_{x_{k+1}}$, and by~(\ref{eq:fs-d}) we obtain that
$(b_k,a_{k+1})\in \delta^B_{x_{k+1}}\circ \varphi^{-1}$, so there exists $b_{k+1}\in B$ such that $(b_k,b_{k+1})\in \delta^B_{x_{k+1}}$ and $(a_{k+1},b_{k+1})\in \varphi $.~Therefore, we have successively built a sequence $b_0,b_1,\ldots ,b_n$ of states of $\cal B$ such that $b_0\in \sigma^B$, $(b_k,b_{k+1})\in \delta^B_{x_{k+1}}$, for every $k$, $0\le k\le n-1$, and $(a_k,b_k)\in \varphi$, for each $k$, $0\le k\le n$.~Moreover, by~(\ref{eq:fs-t}) we obtain that $b_n\in\tau^B$.~Thus, the sequence $b_0,b_1,\ldots ,b_n$ is a successful run of the automaton $\cal B$ on the word $u$ which simulates the original run $a_0,a_1,\ldots ,a_n$ of $\cal A$ on~$u$.~In contrast to forward simulations, where we build the sequence $b_0,b_1,\ldots ,b_n$ moving forward, starting with $b_0$ and ending with $b_n$, in the case of backward simulations we build this sequence moving backward, starting with $b_n$ and ending with $b_0$.~In a similar way we can understand forward and backward simulations between arbitrary fuzzy automata, taking into account degrees of possibility of transitions and degrees of relationship.

In numerous papers dealing with simulations and bisimulations mostly forward simulations and forward~bisimulations have been studied.~They have been usually called just simulations and bisimulations, or {\it strong simulations\/} and {\it strong bisimulations\/} (cf.~\cite{Milner.89,Milner.99,RM-C.00}), and the greatest bisimulation equivalence has been usually called a {\it bisimilarity\/}.~Distinction between forward and backward simulations, and forward and backward bisimulations, has been made, for instance, in \cite{Buch.08,HKPV.98,LV.95} (for various kinds of automata), but~less or more these concepts differ from the concepts having the same name which are considered here.~More~similar to our concepts of forward and backward simulations and bisimulations are those studied in \cite{Brihaye.07}, and in \cite{HMM.07,HMM.09} (for tree automata).~Moreover, backward-forward bisimulations have been discussed in the context of weighted automata in \cite{BLS.05,BLS.06,BP.03,BE.93,Buch.08,EK.01,EM.10}.

Given fuzzy automata ${\cal A}=(A,\delta^A,\sigma^A,\tau^A)$ and ${\cal B}=(B,\delta^B,\sigma^B,\tau^B)$,
a fuzzy relation $\varphi \in {\cal R}(A,B)$ is a~backward simulation between fuzzy automata ${\cal A}$ and ${\cal B}$ if and only if it is a forward simulation between the reverse fuzzy automata $\bar{{\cal A}}$ and $\bar{{\cal B}}$.~Therefore, forward and backward simulations, forward and backward bisimulations,
and backward-forward and forward-backward bisimulations, are mutually dual
concepts, what means that for any statement on some of these concepts
which is universally valid (valid for all fuzzy automata) there is the corresponding universally valid statement on its dual concept.

Now we introduce several other notions and notation that will be used in future work.

For non-empty sets $A$ and $B$ and fuzzy subsets $\eta\in {\cal F}(A)$ and $\xi \in {\cal F}(B)$,
  fuzzy relations $\eta\rightarrow \xi \in {\cal
R}(A,B)$ and $\eta\leftarrow \xi \in {\cal R}(A,B)$ are defined as follows
\begin{align}
\label{eq:rightarrow}
( \eta\rightarrow \xi)(a,b)= (\,\eta (a) \, \rightarrow \, \xi(b)\,) , \\
\label{eq:leftarrow}
(\eta\leftarrow \xi)(a,b)=(\,\xi (b)\, \rightarrow \, \eta(a)\,) ,
\end{align}
for arbitrary $a\in A$ and $b\in B$.~Let us note that $\eta\leftarrow \xi=(\xi\rightarrow \eta)^{-1}$.

We have the following.

\begin{lemma}\label{le:arrows}
Let $A$ and $B$ be non-empty sets and let $\eta\in {\cal F}(A)$ and $\xi \in {\cal F}(B)$.
\begin{itemize}\parskip=0pt
\item[{\rm (a)}] The set of all solutions to the inequality $\eta \circ \chi \le \xi $, where $\chi $ is an unknown fuzzy relation between $A$ and $B$, is the principal ideal of ${\cal R}(A,B)$ generated by the fuzzy relation  $\eta\rightarrow \xi$.
\item[{\rm (b)}] The set of all solutions to the inequality $\chi \circ \xi \le\ \eta $, where $\chi $ is an unknown fuzzy relation between $A$ and $B$, is the principal ideal of ${\cal R}(A,B)$ generated by the fuzzy relation  $\eta\leftarrow \xi$.
\end{itemize}
\end{lemma}

\begin{proof}
These are the well-known results by E. Sanchez (cf.~\cite{San.76,San.77,San.78}).
\end{proof}

It is worth noting that $(\eta\rightarrow \xi )\land (\eta\leftarrow \xi )=\eta\leftrightarrow \xi$, where $\eta\leftrightarrow \xi$ is a fuzzy relation between $A$ and $B$ defined by
\begin{equation}
\label{eq:leftrightarrow}
  (\eta\leftrightarrow \xi)(a,b) =(\, \eta(a) \, \leftrightarrow \,  \xi(b)\,) ,
\end{equation}
for arbitrary $a\in A$ and $b\in B$.

Next, let $A$ and $B$ be non-empty sets and let $\alpha \in {\cal R}(A)$, $\beta\in {\cal R}(B)$ and $\varphi \in {\cal R}(A,B)$.~The {\it right residual\/} of $\varphi $~by $\alpha $ is a fuzzy relation $\varphi /\alpha \in {\cal R}(A,B)$ defined by
\begin{equation}\label{eq:rr.def}
 (\varphi /\alpha)(a,b) = \bigwedge_{a'\in A}\, (\, \alpha (a',a) \rightarrow \varphi(a',b)\, ),
\end{equation}
for all $a\in A$ and $b\in B$, and the {\it left residual\/} of $\varphi $ by $\beta $ is a fuzzy relation $\varphi \backslash\beta \in {\cal R}(A,B)$ defined by
\begin{equation}\label{eq:lr.def}
( \varphi \backslash\beta)(a,b)=  \bigwedge_{b'\in B}\, (\, \beta(b,b') \rightarrow \varphi(a,b')\, ),
\end{equation}
for all $a\in A$ and $b\in B$.~In the case when $A=B$, these two concepts become the well-known concepts of~right and left residuals of fuzzy relations on a set (cf.~\cite{ICB.10}).

We also have the following.

\begin{lemma}\label{le:residuals}
Let $A$ and $B$ be non-empty sets and let $\alpha \in {\cal R}(A)$, $\beta\in {\cal R}(B)$ and $\varphi \in {\cal R}(A,B)$.
\begin{itemize}\parskip=0pt
\item[{\rm (a)}] The set of all solutions to the inequality $\alpha \circ \chi \le \varphi $, where $\chi $ is an unknown fuzzy relation between $A$ and $B$, is the principal ideal of ${\cal R}(A,B)$ generated by the right residual $\varphi /\alpha $ of of $\varphi $ by $\alpha $.
\item[{\rm (b)}] The set of all solutions to the inequality $\chi \circ \beta \le \varphi $, where $\chi $ is an unknown fuzzy relation between $A$ and $B$, is the principal ideal of ${\cal R}(A,B)$ generated by the left residual $\varphi \backslash\beta $ of of $\varphi $ by $\beta $.
\end{itemize}
\end{lemma}

\begin{proof}
These are also results by E. Sanchez (cf.~\cite{San.76,San.77,San.78}).\end{proof}

As we said in the introduction, the problem of deciding whether there is a simulation or bisimulation of a given type between fuzzy automata, and the problem of computing the greatest simulation/bisimulation of this type, will be reduced to the problem of computing the greatest post-fixed point, contained in a given fuzzy relation, of an appropriate isotone function on the lattice of fuzzy relations.~For this purpose we define the following fuzzy relations and functions on the lattice of fuzzy relations.

Let ${\cal A}=(A,\delta^A,\sigma^A,\tau^A)$ and ${\cal B}=(B,\delta^B,\sigma^B,\tau^B)$ be  fuzzy automata.~We define fuzzy relations $\psi^w\in {\cal R}(A,B)$, for $w\in \{fs,bs,fb,bb,fbb,bfb\}$, in the following way:
\begin{align}
&\psi^{fs}= \tau^A\rightarrow \tau^B ,  \label{eq:psi.fs}\\
&\psi^{bs}= \sigma^A\rightarrow \sigma^B , \label{eq:psi.bs}\\
&\psi^{fb}= (\tau^A\rightarrow \tau^B)\land (\tau^A\leftarrow \tau^B)=\tau^A\leftrightarrow \tau^B, \label{eq:psi.fb}\\
&\psi^{bb}= (\sigma^A\rightarrow \sigma^B)\land (\sigma^A\leftarrow \sigma^B)=\sigma^A\leftrightarrow \sigma^B,  \label{eq:psi.bb}\\
&\psi^{fbb}= (\tau^A\rightarrow \tau^B)\land (\sigma^A\leftarrow \sigma^B),
 \label{eq:psi.fbb}\\
&\psi^{bfb}= (\sigma^A\rightarrow \sigma^B)\land (\tau^A\leftarrow \tau^B).\label{eq:psi.bfb}
\end{align}
Moreover, we define functions $\phi^w: {\cal R}(A,B)\to {\cal R}(A,B)$,
for $w\in \{fs,bs,fb,bb,fbb,bfb\}$, as follows:
\begin{align}
&\phi^{fs}(\alpha)=\bigwedge_{x\in X}[(\delta_x^B\circ \alpha^{-1})\backslash \delta_x^A]^{-1} ,\label{eq:phi.fs}\\
&\phi^{bs}(\alpha)=\bigwedge_{x\in X}(\alpha\circ \delta_x^B)/\delta_x^A
, \label{eq:phi.bs}\\
&\phi^{fb}(\alpha)=\bigwedge_{x\in X}[(\delta_x^B\circ \alpha^{-1})\backslash \delta_x^A]^{-1}\land [(\delta_x^A\circ \alpha)\backslash \delta_x^B] = \phi^{fs}(\alpha)\land [\phi^{fs}(\alpha^{-1})]^{-1} ,\label{eq:phi.fb}\\
&\phi^{bb}(\alpha)=\bigwedge_{x\in X}[(\alpha\circ \delta_x^B)/\delta_x^A]\land [(\alpha^{-1}\circ \delta_x^A)/\delta_x^B]^{-1}= \phi^{bs}(\alpha)\land [\phi^{bs}(\alpha^{-1})]^{-1}, \label{eq:phi.bb}\\
&\phi^{fbb}(\alpha)=\bigwedge_{x\in X}[(\delta_x^B\circ \alpha^{-1})\backslash \delta_x^A]^{-1}\land [(\alpha^{-1}\circ \delta_x^A)/\delta_x^B]^{-1} = \phi^{fs}(\alpha)\land[ \phi^{bs}(\alpha^{-1})]^{-1} ,\label{eq:phi.fbb}\\
&\phi^{bfb}(\alpha)=\bigwedge_{x\in X}[(\alpha\circ \delta_x^B)/\delta_x^A]\land [(\delta_x^A\circ \alpha)\backslash \delta_x^B]= \phi^{bs}(\alpha)\land [\phi^{fs}(\alpha^{-1})]^{-1},
\label{eq:phi.bfb}
\end{align}
for any $\alpha\in {\cal R}(A,B)$.~Notice that in the expression ``$\phi^w(\alpha^{-1})$'' ($w\in \{fs,bs\}$) we denote by $\phi^w$  a function from ${\cal R}(B,A)$ into itself.

The next theorem provides equivalent forms of the second and third conditions in the definitions~of~simu\-lations and bisimulations.

\begin{theorem}\label{th:eq2-3}
Let ${\cal A}=(A,\delta^A,\sigma^A,\tau^A)$ and ${\cal B}=(B,\delta^B,\sigma^B,\tau^B)$ be fuzzy automata and
$w\in \{fs,bs,fb,bb,fbb,bfb\}$. A fuzzy relation $\varphi\in {\cal R}(A,B)$ satisfies conditions $(w$-$2)$ and $(w$-$3)$ if and only if it satisfies
\begin{equation}\label{eq:eq2-3}
\varphi \le \phi^w(\varphi), \qquad \varphi\le \psi^w .
\end{equation}
\end{theorem}

\begin{proof}
We will prove only the case $w=fs$.~The assertion concerning the case
$w=bs$ follows by the duality, and according to equations (\ref{eq:psi.fb})--(\ref{eq:psi.bfb})
and (\ref{eq:phi.fb})--(\ref{eq:phi.bfb}),
all other assertions can be obtained by
the first two.

Consider an arbitrary $\varphi\in {\cal R}(A, B)$.~According to Lemma \ref{le:arrows}(b), $\varphi $ satisfies condition $(fs$-$3)$ if and only if $\varphi^{-1}\le \tau^{B}\leftarrow\tau^{A}=
(\tau^{A}\rightarrow\tau^{B})^{-1}$, which is equivalent to $\varphi\le\tau^A\rightarrow\tau^{B}=\psi^{fs}$.
Therefore, $\varphi$ satisfies $(fs$-$3)$ if and only if $\varphi\le\psi^{fs}$.~

On the other hand, $\varphi $ satisfies $(fs$-$2)$ if and only if
\[
\varphi^{-1}(b,a)\otimes \delta_x^A(a,a')\le (\delta_x^B\circ \varphi^{-1})(b,a'),
\]
for all $a,a'\in A$, $b\in B$ and $x\in X$.~According to the adjunction property,
this is equivalent to
\[
\varphi^{-1}(b,a)\le\bigwedge_{a'\in A}[\delta_{x}^{A}(a,a')\rightarrow (\delta_x^B\circ \varphi^{-1}(b,a'))]=((\delta_x^B\circ \varphi^{-1})\backslash \delta_x^A)(b,a)
\]
 for all $a\in A$, $b\in B$ and $x\in X$, which is further equivalent to
\[
\varphi(a,b)\le\bigwedge_{x\in X}[(\delta_x^B\circ \varphi^{-1})\backslash \delta_x^A]^{-1}(a,b)=(\phi^{fs}(\varphi))(a,b)
\]
for all $a\in A$ and $b\in B$.~Therefore, $\varphi $ satisfies $(fs$-$2)$ if and only if $\varphi\le \phi^{fs}(\varphi)$.

Now we conclude that a fuzzy relation $\varphi\in {\cal R}(A, B)$ satisfies $(fs$-$2)$ and $(fs$-$3)$ if and only if it satisfies~(\ref{eq:eq2-3}) (for
$w=fs)$, which was to be proved.
\end{proof}

\section{Computation of the greatest simulations and bisimulations}\label{sec:main}

We will provide a method for computing the greatest simulations and bisimulations
between fuzzy~auto\-mata adapting the method developed in \cite{ICB.10} for
computing the greatest post-fixed points of an isotone function on the lattice
of fuzzy relations on a single set.

Let $A$ and $B$ be non-empty sets  and let $\phi :{\cal R}(A,B)\to {\cal R}(A,B)$ be an isotone~function, i.e., let  $\alpha\le \beta $ implies $\phi(\alpha)\le \phi (\beta)$, for all $\alpha ,\beta\in {\cal R}(A,B)$.~A fuzzy relation $\alpha\in {\cal R}(A,B)$ is called a {\it post-fixed~point\/} of $\phi $ if $\alpha\le \phi(\alpha)$.~The well-known Knaster-Tarski fixed point theorem (stated and proved in a more general context, for complete lattices) asserts that the set of all post-fixed points of $\phi $ form a
complete lattice (cf.~\cite{Rom.08}).~Moreover, for any fuzzy relation $\psi \in {\cal R}(A,B)$ we have that the set of all post-fixed points of $\phi
$ contained in $\psi $ is also a complete lattice.~According to Theorem \ref{th:eq2-3}, our main task is to find an effective procedure for computing the greatest post-fixed point of the function $\phi^w$ contained in the fuzzy relation
$\psi^w$, for each $w\in \{fs,bs,fb,bb,fbf,bfb\}$.

It should be noted that the set of all post-fixed points of an isotone function
on a complete lattice is always non-empty, because it contains the least element
of this complete lattice.~However, this set may~consist only of that single
element.~In our case, when we are working with a lattice of fuzzy relations,
the empty relation may be the only post-fixed point, whereas we have defined
simulations and bisimulations to be non-empty fuzzy relations.~This requirement is necessary because the empty
relation can not satisfy the condition ($w$-1), unless the fuzzy set of initial states or the fuzzy set of terminal states is also empty.~Therefore, our task is actually to find an effective procedure for deciding whether there is a non-empty post-fixed point of $\phi^w$ contained in $\psi^w$, and if it exists, then find the greatest one.

Let $\phi :{\cal R}(A,B)\to {\cal R}(A,B)$ be an isotone~function and
$\psi \in {\cal R}(A,B)$. We define a sequence $\{\varphi_k\}_{k\in \mathbb{N}}$ of fuzzy relations from ${\cal R}(A,B)$ by
\begin{equation}\label{eq:seq}
\varphi_1=\psi, \qquad \varphi_{k+1}=\varphi_k\land \phi(\varphi_k), \ \ \text{for each}\ k\in \mathbb{N} .
\end{equation}
The sequence  $\{\varphi_k\}_{k\in \mathbb{N}}$ is obviously descending.~If we
denote by $\hat\varphi
$ the greatest post-fixed point of $\phi$ contained in $\psi$, we
can easily verify that
\begin{equation}\label{eq:hat.phi}
\hat\varphi \le \bigwedge_{k\in \mathbb{N}} \varphi_k .
\end{equation}
Now two very important questions arise.~First, under what conditions the equality holds in (\ref{eq:hat.phi})?~Even more important question is: under what conditions the
sequence  $\{\varphi_k\}_{k\in \mathbb{N}}$ is finite? If this sequence is finite,
then it is not hard to show that there exists $k\in \mathbb{N}$ such that $\varphi_k=\varphi_m$, for every $m\ge k$, i.e., there exists $k\in \mathbb{N}$ such that the sequence stabilizes on $\varphi_k$.~We can recognize that the sequence has stabilized when we find the smallest $k\in \mathbb{N}$ such that $\varphi_k=\varphi_{k+1}$.
In this case  $\hat\varphi = \varphi_k$, and we have an algorithm which computes
$\hat\varphi$ in a finite number of steps.

Some conditions under which equality holds in (\ref{eq:hat.phi}) or the sequence
is finite were found in \cite{ICB.10}, for fuzzy relations on a single set.~It is easy to show that the same results are also valid for fuzzy relations between two sets.~In the sequel we present these results.

First we note that a sequence $\{\varphi_k\}_{k\in \mathbb{N}}$ of fuzzy relations from ${\cal R}(A,B)$ is finite if and only it it is {\it image-finite\/}, which means that the set $\bigcup_{k\in \mathbb{N}}\im (\varphi_k)$ is finite.~Next, the function $\phi :{\cal R}(A,B)\to {\cal R}(A,B)$ is called {\it image-localized\/} if there exists a finite $K\subseteq L$ such that for~each fuzzy~relation $\varphi\in {\cal R}(A,B)$ we have
\begin{equation}\label{eq:im.loc}
\im (\phi(\varphi))\subseteq \langle K\cup \im (\varphi)\rangle ,
\end{equation}
where $\langle K\cup \im (\varphi)\rangle$ denotes the subalgebra of $\cal L$ generated by the set $K\cup \im (\varphi)$.~Such $K$ will be called a {\it localization set\/} of the function $\phi $.

Theorem analogous to the following theorem was proved in \cite{ICB.10} for fuzzy relations on a single set, but~its proof can be easily transformed into the proof of the corresponding theorem concerning fuzzy relations between two sets.

\begin{theorem}\label{th:phi.loc}
Let the function $\phi $ be image-localized, let $K$ be its localization set, let $\psi \in {\cal R}(A,B)$, and let $\{\varphi_k\}_{k\in \mathbb{N}}$ be a sequence of fuzzy relations in ${\cal R}(A,B)$ defined by $(\ref{eq:seq})$. Then
\begin{equation}\label{eq:UimRk}
\bigcup_{k\in \mathbb{N}}\im (\varphi_k) \subseteq \langle K\cup \im (\psi)\rangle .
\end{equation}
If, moreover, $\langle K\cup \im (\psi)\rangle$ is a finite subalgebra of $\cal L$, then the sequence
$\{\varphi_k\}_{k\in \mathbb{N}}$ is finite.
\end{theorem}

Going back now to the functions $\phi^w$, for $w\in \{fs,bs,fb,bb,fbb,bfb\}$, we prove the following.

\begin{theorem}\label{th:phiw}
Let ${\cal A}=(A,\delta^A,\sigma^A,\tau^A)$ and ${\cal B}=(B,\delta^B,\sigma^B,\tau^B)$
be arbitrary automata.

For any $w\in \{fs,bs,fb,bb,fbb,bfb\}$ the function $\phi^w$ is isotone and image-localized.
\end{theorem}

\begin{proof}
We will prove only the case $w=fs$.~The assertion concerning the case
$w=bs$ follows by the duality, and according to equations (\ref{eq:psi.fb})--(\ref{eq:psi.bfb})
and (\ref{eq:phi.fb})--(\ref{eq:phi.bfb}), all other assertions can be derived from the first two.

Let $\varphi_1,\varphi_2\in {\cal R}(A, B)$ be fuzzy relations such that $\varphi_1\le \varphi_2$, and consider the following systems of fuzzy relation inequalities:
\begin{align}
 \chi^{-1}\circ \delta_x^A\le \delta_x^B\circ \varphi_1^{-1},\hspace{0.1 in} x\in X;\label{lem.proof.varphi} \\
\chi^{-1}\circ \delta_x^A\le \delta_x^B\circ \varphi_2^{-1},\hspace{0.1 in} x\in X.\label{lem.proof.psi}
\end{align}
where $\chi\in {\cal R}(A,B)$ is an unknown fuzzy relation.~Using Lemma \ref{le:residuals} (b) and the definition of an inverse~rela\-tion, it can be easily shown that the set of all solutions to system (\ref{lem.proof.varphi}) (resp. (\ref{lem.proof.psi})) form a principal ideal~of ${\cal R}(A,B)$ generated by $\phi^{fs}(\varphi_1)$ (resp. $\phi^{fs}(\varphi_2)$).~Since for each $x\in X$ we have that $\delta_x^B\circ \varphi_1^{-1}\le\delta_x^B\circ \varphi_2^{-1}$,~we~conclude that every solution to (\ref{lem.proof.varphi}) is a solution to (\ref{lem.proof.psi}).~Consequently,
 $\phi^{fs}(\varphi_1)$ is a solution to (\ref{lem.proof.psi}), so  $\phi^{fs}(\varphi_1)\le\phi^{fs}(\varphi_2)$.
Therefore, we have proved that $\phi^{fs}$ is an isotone function.

Next, let $K=\bigcup_{x\in X}(\im(\delta_{x}^{A})\cup \im(\delta_{x}^{B}))$
and let $\varphi\in {\cal R}(A,B)$ be an arbitrary fuzzy relation.~It is
evident that $\im(\phi^{fs}(\varphi))\subseteq\langle K\bigcup \im(\varphi)\rangle$,
and since the input alphabet $X$ is finite, then $K$ is also finite.~This
confirms that the function $\phi^{fs}$ is image-localized.
\end{proof}

Now we are ready to prove the main result of this paper, which provides algorithms for deciding~whether there is a simulation or bisimulation of a given type between fuzzy automata, and for computing the greatest simulations and bisimulations, when they exist.

\begin{theorem}\label{th:alg}
Let ${\cal A}=(A,\delta^A,\sigma^A,\tau^A)$ and ${\cal B}=(B,\delta^B,\sigma^B,\tau^B)$ be fuzzy automata, let
$w\in \{fs,bs,fb,bb,fbb,bfb\}$, and let a sequence $\{\varphi_k\}_{k\in \mathbb{N}}$ of fuzzy relations from ${\cal R}(A,B)$ be defined by
\begin{equation}\label{eq:alg.seq}
\varphi_1=\psi^w, \qquad \varphi_{k+1}=\varphi_k\land \phi^w(\varphi_k), \ \ \text{for each}\ k\in \mathbb{N} .
\end{equation}
If $\langle \im (\psi^w)\cup \bigcup_{x\in X}(\,\im (\delta^A_x)\cup \im(\delta^B_x)\,)\rangle $ is a finite subalgebra of $\cal L$, then the following is true:
\begin{itemize}\parskip0pt
\item[{\rm (a)}] the sequence $\{\varphi_k\}_{k\in \mathbb{N}}$ is finite and
descending, and there is the least natural number $k$ such that $\varphi_k=\varphi_{k+1}$;
\item[{\rm (b)}] $\varphi_k$ is the greatest fuzzy relation in ${\cal R}(A,B)$ which satisfies $(w\text{-}2)$ and $(w\text{-}3)$;
\item[{\rm (c)}] if $\varphi_k$ satisfies $(w\text{-}1)$, then it is the greatest fuzzy relation in ${\cal R}(A,B)$ which satisfies $(w\text{-}1)$, $(w\text{-}2)$ and $(w\text{-}3)$;
\item[{\rm (d)}] if $\varphi_k$ does not satisfy $(w\text{-}1)$, then there is no any fuzzy relation in ${\cal R}(A,B)$ which satisfies $(w\text{-}1)$, $(w\text{-}2)$ and $(w\text{-}3)$.
\end{itemize}
\end{theorem}

\begin{proof}
We will prove only the case $w=fs$.~All other cases can be proved in a similar
manner.

So, let $\langle \im (\psi^{fs})\cup \bigcup_{x\in X}(\,\im (\delta^A_x)\cup \im(\delta^A_x)\,)\rangle $ be a finite subalgebra of $\cal L$.

(a) According to Theorems \ref{th:phiw} and \ref{th:phi.loc}, the sequence $\{\varphi_{k}\}_{k\in N}$ is finite and
descending, so there are~$k,m\in \mathbb{N}$ such that $\varphi_k=\varphi_{k+m}$,
whence $\varphi_{k+1}\le \varphi_k=\varphi_{k+m}\le \varphi_{k+1}$.~Thus, there is $k\in \mathbb{N}$ such that $\varphi_k=\varphi_{k+1}$,~and conse\-quently,
there is the least natural number having this property.

(b) By $\varphi_k=\varphi_{k+1}=\varphi_k\land \phi^{fs}(\varphi_{k})$ we
obtain that $\varphi_k\le \phi^{fs}(\varphi_{k})$, and also, $\varphi_k\le
\varphi_1=\psi^{fs}$.~Therefore,~by~Theo\-rem \ref{th:eq2-3} it follows that
$\varphi_k$ satisfies $(fs\text{-}2)$ and $(fs\text{-}3)$.

Let $\varphi \in{\cal R}(A,B)$ be an arbitrary fuzzy relation which satisfies
$(fs\text{-}2)$ and $(fs\text{-}3)$.~As we have already noted, $\varphi$ satisfies $(fs\text{-}3)$ if and only if $\varphi\le\psi^{fs}=\varphi_{1}$.~Next, suppose that $\varphi \le \varphi_n$, for some $n\in \mathbb{N}$.~Then for every $x\in X$ we have that $\varphi^{-1}\circ \delta_x^A\le \delta_x^B\circ \varphi^{-1}\le  \delta_x^B\circ \varphi_n^{-1}$, and according to Lemma \ref{le:residuals} (b), $\varphi^{-1}\le (\delta_x^B\circ\varphi_n^{-1})\setminus \delta_x^A$, i.e., $\varphi \le [(\delta_x^B\circ\varphi_n^{-1})\setminus \delta_x^A]^{-1}=\phi^{fs}(\varphi_n)$.~Therefore,
$\varphi\le \varphi_n\land \phi^{fs}(\varphi_{n})= \varphi_{n+1}$.~Now, by induction we obtain  that $\varphi \le \varphi_n$, for every $n\in \mathbb{N}$, and hence, $\varphi \le \varphi_{k} $.~This means that $\varphi_{k} $~is~the~greatest fuzzy relation in ${\cal R}(A,B)$ satisfying $(fs\text{-}2)$ and $(fs\text{-}3)$.

(c) This follows immediately from (b).

(d) Suppose that   $\varphi_{k}$ does not satisfy $(fs\text{-}1)$.~Let $\varphi \in{\cal R}(A,B)$ be an arbitrary fuzzy relation which satisfies $(fs\text{-}1)$,
$(fs\text{-}2)$ and $(fs\text{-}3)$.~According to (b) of this theorem, $\varphi\le \varphi_{k}$, so we  have that $\sigma^{A}\le \sigma^{B}\circ\varphi^{-1}\le\sigma^{B}\circ\varphi_k^{-1}$.~But, this contradicts our starting assumption
that   $\varphi_{k}$ does not satisfy $(fs\text{-}1)$.~Hence, we conclude that there is no any fuzzy relation in ${\cal R}(A,B)$ which satisfies $(fs\text{-}1)$, $(fs\text{-}2)$ and $(fs\text{-}3)$.
\end{proof}

Next, we will consider the case when ${\cal L}=(L,\land ,\lor , \otimes ,\to , 0,
1)$ is a complete residuated lattice satisfying
the following conditions:
\begin{eqnarray}\label{eq:infd}
x\lor \bigl(\bigwedge_{i\in I}y_i\bigr) = \bigwedge_{i\in I}(x\lor y_i) ,\\
\label{eq:infdm}
x\otimes \bigl(\bigwedge_{i\in I}y_i\bigr) = \bigwedge_{i\in I}(x\otimes y_i) ,
\end{eqnarray}
for all $x\in L$ and $\{y_i\}_{i\in I}\subseteq L$.~Let us note that if ${\cal L}=([0,1],\land ,\lor , \otimes ,\to , 0, 1)$, where $[0,1]$ is the real unit interval
and $\otimes $ is a left-con\-tin\-uous t-norm on $[0,1]$, then (\ref{eq:infd}) follows
immediately by linearity of $\cal L$, and $\cal L$ satisfies (\ref{eq:infdm}) if and only if $\otimes $ is
a continuous t-norm, i.e., if and only if $\cal L$ is a $BL$-algebra~(cf.~\cite{Bel.02,BV.05}).~Therefore,~conditions
(\ref{eq:infd}) and (\ref{eq:infdm}) hold for~every $BL$-algebra on the real unit interval.~In particular,
the {\L}ukasiewicz, Goguen (product)~and~G\"odel structures fulfill {\rm (\ref{eq:infd})} and {\rm (\ref{eq:infdm})}.

Under these conditions we have the following.

\begin{theorem}\label{th:inf}
Let ${\cal A}=(A,\delta^A,\sigma^A,\tau^A)$ and ${\cal B}=(B,\delta^B,\sigma^B,\tau^B)$ be fuzzy automata, let
$w\in \{fs,bs,fb,bb,fbb,bfb\}$, let $\{\varphi_k\}_{k\in \mathbb{N}}$ be a sequence of fuzzy relations from ${\cal R}(A,B)$ defined by $(\ref{eq:alg.seq})$, and let
\begin{equation}\label{eq:inf}
{\varphi} = \bigwedge_{k\in \mathbb{N}}\varphi_k .
\end{equation}
If $\cal L$ is a complete residuated lattice satisfying $(\ref{eq:infd})$ and $(\ref{eq:infdm})$, then the following is true:
\begin{itemize}\parskip0pt
\item[{\rm (a)}] ${\varphi} $ is the greatest fuzzy relation in ${\cal R}(A,B)$ which satisfies $(w\text{-}2)$ and $(w\text{-}3)$;
\item[{\rm (b)}] if ${\varphi} $ satisfies $(w\text{-}1)$, then it is the greatest fuzzy relation in ${\cal R}(A,B)$ which satisfies $(w\text{-}1)$, $(w\text{-}2)$ and $(w\text{-}3)$;
\item[{\rm (c)}] if ${\varphi} $ does not satisfy $(w\text{-}1)$, then there is no any fuzzy relation in ${\cal R}(A,B)$ which satisfies $(w\text{-}1)$, $(w\text{-}2)$ and $(w\text{-}3)$.
\end{itemize}
\end{theorem}

\begin{proof}
We will prove only the case $w=fs$.~All other cases can be proved similarly.

So, let $\cal L$ be a complete residuated lattice satisfying $(\ref{eq:infd})$ and $(\ref{eq:infdm})$.~First, notice that it has been  shown in \cite{CSIP.10} that if {\rm (\ref{eq:infd})} holds, then for all descending sequences
$\{x_k\}_{k\in \mathbb{N}}, \{y_k\}_{k\in \mathbb{N}}\subseteq L$ we have
\begin{equation}\label{eq:desc.seq}
\bigwedge_{k\in \mathbb{N}}(x_k\lor y_k) = \bigl(\bigwedge_{k\in \mathbb{N}}x_k\bigr)\lor
\bigl(\bigwedge_{k\in \mathbb{N}}y_k\bigr).
\end{equation}

(a) For arbitrary $x\in X$, $a\in A$ and $b\in B$ we have that
\[
\begin{aligned}
\Biggl(\,\bigwedge_{k\in \mathbb{N}}(\delta_x^B\circ \varphi_k^{-1})\Biggr)\,(b,a)&=\bigwedge_{k\in \mathbb{N}}(\delta_x^B\circ \varphi_k^{-1})(b,a)=\bigwedge_{k\in \mathbb{N}}\Biggl(\,\bigvee_{b'\in
B}\delta_x^B(b,b')\otimes \varphi_k^{-1}(b',a)\Biggr) && \\
&= \bigvee_{b'\in B}\Biggl(\,\bigwedge_{k\in \mathbb{N}}\delta_x^B(b,b')\otimes \varphi_k^{-1}(b',a)\Biggr) && \text{(by (\ref{eq:desc.seq}))} \\
&= \bigvee_{b'\in B}\Biggl(\,\delta_x^B(b,b')\otimes\biggl( \bigwedge_{k\in \mathbb{N}}\varphi_k^{-1}(b',a)\biggr)\Biggr) && \text{(by (\ref{eq:infdm}))} \\
&= \bigvee_{b'\in B}\biggl(\,\delta_x^B(b,b')\otimes \varphi^{-1}(b',a)\biggr)
=(\delta_x^B\circ \varphi^{-1})(b,a),  &&
\end{aligned}
\]
which means that
\[
\bigwedge_{k\in \mathbb{N}}\delta_x^B\circ \varphi_k^{-1}=\delta_x^B\circ \varphi^{-1},
\]
for every $x\in X$.~The use of condition (\ref{eq:desc.seq}) is justified by the facts that $B$ is finite, and that
$\{\varphi_k^{-1}(b',a)\}_{k\in \mathbb{N}}$ is a descending sequence, so
$\{\delta_x^B(b,b')\otimes \varphi_k^{-1}(b',a)\}_{k\in \mathbb{N}}$ is also a descending sequence.

Now, for all $x\in X$ and $k\in \mathbb{N}$ we have that
\[
\varphi\le \varphi_{k+1}\le \phi^{fs}(\varphi_k)=[(\delta_x^B\circ \varphi_k^{-1})\backslash
\delta_x^A]^{-1},
\]
which is equivalent to
\[
\varphi^{-1}\circ \delta_x^A\le \delta_x^B\circ \varphi_k^{-1}.
\]
As the last inequality holds for every $k\in \mathbb{N}$, we have that
\[
\varphi^{-1}\circ \delta_x^A\le \bigwedge_{k\in \mathbb{N}}\delta_x^B\circ \varphi_k^{-1}=\delta_x^B\circ \varphi^{-1},
\]
for every $x\in X$.~Therefore, $\varphi $ satisfies ($fs$-2).~Moreover, $\varphi
\le \varphi_1=\psi^{fs}$, so $\varphi $ also satisfies ($fs$-3).

Next, let $\alpha\in {\cal R}(A,B)$ be an arbitrary fuzzy relation satisfying $(fs\text{-}2)$ and $(fs\text{-}3)$.~According to Theorem~\ref{th:eq2-3},
$\alpha \le \phi^{fs}(\alpha )$ and $\alpha \le \psi^{fs}=\varphi_1$.~By induction we can easily prove that
$\alpha\le\varphi_{k}$ for every $k\in \mathbb{N}$, and therefore, $\alpha\le
\varphi$.~This means that $\varphi$ is the greatest fuzzy relation in ${\cal R}(A,B)$ which satisfies $(fs\text{-}2)$ and $(fs\text{-}3)$.

The assertion (b) follows immediately from (a), whereas the assertion (c) can be proved in the same way as the assertion (d) of  Theorem \ref{th:alg}.
\end{proof}

In some situations we do not need simulations and bisimulations that are fuzzy relations, but those~that are ordinary crisp relations.~Moreover, in cases where our algorithms for computing the greatest simulations and bisimulations fail to terminate in a finite number of steps, we can search for the greatest crisp simulations and bisimulations.~They can be understood as a kind of ``approximations'' of the greatest~fuzzy simulations and bisimulations.~Here we show that the above given algorithms for computing the greatest fuzzy simulations and bisimulations can be modified to compute the greatest crisp simulations and bisimulations.~The new algorithms terminate in a finite number of~steps, independently of the properties of the underlying structure of truth values.~In the next section we will give an example which~shows~that~the~great\-est crisp simulations and bisimulations~can not be obtained simply by taking the crisp parts of the greatest fuzzy simulations and bisimulations.~In fact, our example shows that there may be a fuzzy simulation/bisimulation of a given type between two fuzzy automata, but there is not any crisp simulation/bisimulation of this type between them.

Let $A$ and $B$ be non-empty finite sets, and let ${\cal R}^{c}(A,B)$ denote the set of all crisp relations from ${\cal R}(A,B)$.~It is not hard to verify that ${\cal R}^{c}(A,B)$ is a complete sublattice of ${\cal R}(A,B)$, i.e., the meet and the join in ${\cal R}(A,B)$ of an arbitrary family
of crisp relations from ${\cal R}^{c}(A,B)$ are also crisp relations (in
fact, they coincide with~the~ordi\-nary intersection and union of crisp relations).~Moreover,
for each fuzzy relation  $\varphi\in {\cal R}(A,B)$ we have that $\varphi^{c}\in {\cal R}^{c}(A,B)$, where $\varphi^{c}$ denotes the \textit{crisp part\/} of a fuzzy relation $\varphi $ (in some sources called the \textit{kernel\/} of $\varphi $), i.e., a function $\varphi^{c}:A\times B\to \{0,1\}$ defined
by $\varphi^{c}(a,b)=1$, if $\varphi(a,b)=1$, and $\varphi^{c}(a,b)=0$, if
$\varphi(a,b)<1$, for arbitrary $a\in A$ and $b\in B$.~Equivalently,
$\varphi^{c}$ is considered as an ordinary crisp relation between $A$ and $B$~given
by $\varphi^{c}=\{(a,b)\in A\times B\mid \varphi(a,b)=1\}$.

For each function $\phi :{\cal R}(A,B)\to {\cal R}(A,B)$ we define a function $\phi^{c}: {\cal R}^c(A,B)\to {\cal R}^c(A,B)$~by
\[
\phi^{\mathrm{c}}(\varphi)=(\phi(\varphi))^{\mathrm{c}}, \ \ \text{for any}\ \varphi\in {\cal R}^c(A,B).
\]
If $\phi $ is isotone, then it can be easily shown that $\phi^{\mathrm{c}}$ is also an isotone function.

We have that the following is true.

\begin{proposition}\label{prop:crisp.alg}
Let $A$ and $B$ be non-empty finite sets, let\ $\phi :{\cal R}(A,B)\to {\cal R}(A,B)$ be an isotone function and let  $\psi\in {\cal R}(A,B)$ be a given fuzzy relation.~A~crisp relation $\varrho\in {\cal R}^c(A,B)$ is the greatest  crisp solution in ${\cal R}(A,B)$ to the system
\begin{equation}\label{eq:c.fri}
\chi\le \phi (\chi), \qquad\qquad \chi\le \psi,
\end{equation}
 if and only if it is the greatest solution in ${\cal R}^c(A,B)$ to the system
\begin{equation}\label{eq:c.cri}
\xi\le \phi^{c} (\xi), \qquad\qquad \xi\le \psi^c,
\end{equation}
 where $\chi$ is an unknown fuzzy relation and $\xi $ is an unknown crisp relation.

Furthermore, a sequence $\{\varrho_k\}_{k\in \mathbb{N}}\subseteq {\cal R}(A,B)$ defined by
\begin{equation}\label{eq:c.seq}
\varrho_1=\psi^c, \ \ \varrho_{k+1}=\varrho_k\land \phi^{c}(\varrho_k), \ \ \text{for every}\ k\in \mathbb{N},
\end{equation}
is a finite descending sequence of crisp relations, and the least
member of this sequence is the greatest solution to the system $(\ref{eq:c.cri})$ in ${\cal R}^c(A,B)$.
\end{proposition}

\begin{proof}
The proof of this proposition can be obtained simply by translating the proof of Theorem 5.8. \cite{ICB.10} to the case of relations between the two sets.
\end{proof}

Taking $\phi $ to be any of the functions $\phi^w$, for $w\in \{fs,bs,fb,bb,fbb,bfb\}$, Proposition \ref{prop:crisp.alg} gives algorithms for deciding whether there is a crisp simulation or bisimulation of a given type between fuzzy automata,~and for computing the greatest crisp simulations and bisimulations, when they exist.~As we have seen~in Proposi\-tion~\ref{prop:crisp.alg}, these algorithms always terminate in a finite number of steps, independently of the proper\-ties of the underlying structure of truth values.~However, as we have already mentioned, in the next section we will give an example which shows that there may be a fuzzy simulation/bisimulation of a given type~between two fuzzy automata, and there is not any crisp simulation/bisimulation of this type between them.

It is worth noting that functions $(\phi^w)^c$, for all $w\in \{fs,bs,fb,bb,fbb,bfb\}$, can be characterized as follows:
\[
\begin{aligned}
&(a,b)\in (\phi^{fs})^{c}(\varrho )\ \iff\ \ (\forall x\in X)(\forall a'\in A)\,\delta_x^A(a,a')\le (\delta_x^B\circ \varrho^{-1})(b,a'), \\
&(a,b)\in (\phi^{bs})^{c}(\varrho )\ \iff\ \ (\forall x\in X)(\forall a'\in A)\,\delta_x^A(a',a)\le (\varrho\circ\delta_x^B)(a',b), \\
&(\phi^{fb})^{c}(\varrho )=(\phi^{fs})^{c}(\varrho )\land [(\phi^{fs})^{c}(\varrho^{-1})]^{-1},\ \
(\phi^{bb})^{c}(\varrho )=(\phi^{bs})^{c}(\varrho )\land [(\phi^{bs})^{c}(\varrho^{-1})]^{-1}, \\
&(\phi^{fbb})^{c}(\varrho )=(\phi^{fs})^{c}(\varrho )\land [(\phi^{bs})^{c}(\varrho^{-1})]^{-1},\ \
(\phi^{bfb})^{c}(\varrho )=(\phi^{bs})^{c}(\varrho )\land [(\phi^{fs})^{c}(\varrho^{-1})]^{-1},
\end{aligned}
\]
for all $\varrho\in {\cal R}^c(A,B)$, $a\in A$ and $b\in B$.

\section{Computational examples}\label{sec:ex}

In this section we give examples which demonstrate the application of our algorithms and clarify~relationships between different types of simulations and bisimulations.

The first example demonstrates the case when there are all types of simulations and bisimulations, but there is not any crisp bisimulation between two given automata.

\begin{example}\label{ex:first}\rm
Let ${\cal A}=(A,\delta^A,\sigma^A,\tau^A)$ and ${\cal B}=(B,\delta^B,\sigma^B,\tau^B)$ be automata over the G\"odel  structure and an alphabet $X=\{x,y\}$, with $|A|=3$,  $|B|=2$, and fuzzy transition relations and fuzzy sets of
initial and terminal states which are represented by the following fuzzy matrices and vectors:
\begin{align}\label{eq:faA}
&\sigma^A=\begin{bmatrix} 1 & 1 & 1 \end{bmatrix},
&&\delta_x^A=\begin{bmatrix}
 1  & 0.3 & 0.4 \\
0.5 &  1  & 0.3 \\
0.4 & 0.6 & 0.7
\end{bmatrix},
&&\delta_y^A=\begin{bmatrix}
0.5 & 0.6 & 0.2 \\
0.3 & 0.3 & 0.4 \\
0.7 & 0.7 &  1
\end{bmatrix},
&&\tau^A= \begin{bmatrix}
1 \\ 1 \\ 1
\end{bmatrix}, &&\\ \label{eq:faB}
&\sigma^B=\begin{bmatrix} 1 & 1 \end{bmatrix},
&&\delta_x^B=\begin{bmatrix}
 1  & 0.6 \\
0.6 & 0.7
\end{bmatrix},
&&\delta_y^B=\begin{bmatrix}
0.6 & 0.6 \\
0.7 &  1
\end{bmatrix},
&&\tau^B= \begin{bmatrix}
1 \\ 1
\end{bmatrix}. &&
\end{align}
Using algorithms based on Theorem \ref{th:alg} we obtain that there are all types of simulations and bisimulations
between fuzzy automata $\cal A$ and $\cal B$, and the greatest simulations and bisimulations are given by matrices
\[\small
\varphi^{fs}=\begin{bmatrix}
 1  & 0.7 \\
 1  & 0.7 \\
0.6 &  1
\end{bmatrix}, \ \
\varphi^{bs}=\begin{bmatrix}
 1  & 0.7 \\
 1  & 0.7 \\
0.7 &  1
\end{bmatrix}, \ \
\varphi^{fb}=\begin{bmatrix}
 1  & 0.6 \\
 1  & 0.6 \\
0.6 &  1
\end{bmatrix}, \ \
\varphi^{bb}=\begin{bmatrix}
 1  & 0.7 \\
 1  & 0.7 \\
0.7 &  1
\end{bmatrix}, \ \
\varphi^{fbb}=\begin{bmatrix}
 1  & 0.7 \\
 1  & 0.7 \\
0.6 &  1
\end{bmatrix}, \ \
\varphi^{bfb}=\begin{bmatrix}
 1  & 0.6 \\
 1  & 0.6 \\
0.7 &  1
\end{bmatrix}.
\]
On the other hand, using the version of the algorithms for crisp simulations and bisimulations, we obtain that
there is not any crisp bisimulation between fuzzy automata $\cal A$ and $\cal B$.
\end{example}

The second example demonstrates the case when there are is a forward bisimulation, but there is not any other type
of bisimulations between two given automata.

\begin{example}\rm
Let us change $\sigma^A$ in (\ref{eq:faA}) and $\sigma^B$ in (\ref{eq:faB}) to
\[
\sigma^A=\begin{bmatrix} 0 & 1 & 0 \end{bmatrix},\qquad \sigma^B=\begin{bmatrix} 1 & 0.5 \end{bmatrix}.
\]
Then the greatest forward bisimulation between fuzzy automata $\cal A$ and $\cal B$ is given by
\[
\varphi^{fb}=\begin{bmatrix}
 1  & 0.6 \\
 1  & 0.6 \\
0.6 &  1
\end{bmatrix},
\]
but there is not any other type of bisimulations between $\cal A$ and $\cal B$.
\end{example}

Due to the duality between forward and backward bisimulations, it is possible to construct automata between which there is a backward bisimulation, but there is not any other type of bisimulations.

We can also give an example which demonstrates the case when there are is a backward-forward~bisimulation, but there is not any other type of bisimulations between two given automata.

\begin{example}\rm
Let us change $\sigma^A$ in (\ref{eq:faA}) and $\sigma^B$ in (\ref{eq:faB}) to
\[
\sigma^A=\begin{bmatrix} 0 & 0 & 1 \end{bmatrix},\qquad \sigma^B=\begin{bmatrix} 0.7 & 1 \end{bmatrix}.
\]
Then the greatest backward-forward bisimulation between fuzzy automata $\cal A$ and $\cal B$ is given by
\[
\varphi^{bfb}=\begin{bmatrix}
 1  & 0.6 \\
 1  & 0.6 \\
0.7 &  1
\end{bmatrix},
\]
but there is not any other type of bisimulations between $\cal A$ and $\cal B$.
\end{example}

Due to the duality between backward-forward and forward-backward bisimulations, it is possible~to~con\-struct automata between which there is a backward bisimulation, but there is not any other type of bisimulations.

Next, we give an example which demonstrates the case when there is not any type of simulations and bisimulations between two given automata.

\begin{example}\rm
Let us change $\sigma^A$ in (\ref{eq:faA}) and $\sigma^B$ in (\ref{eq:faB}) to
\[
\sigma^A=\begin{bmatrix} 1 & 0 & 0 \end{bmatrix},\qquad \sigma^B=\begin{bmatrix} 0.5 & 1 \end{bmatrix}.
\]
Then there is not any type of simulations and bisimulations between $\cal A$ and $\cal B$.
\end{example}

The following example considers the case (over the product structure) when the sequence of fuzzy~relations defined by (\ref{eq:alg.seq}) is infinite, and its intersection is the greatest forward bisimulation between the given fuzzy automata.

\begin{example}\label{ex:inf.seq}\rm
Let ${\cal A}=(A,\delta^A,\sigma^A,\tau^A)$ and ${\cal B}=(B,\delta^B,\sigma^B,\tau^B)$ be automata over the Goguen (product) structure and an alphabet $X=\{x\}$, with $|A|=3$,  $|B|=2$, and fuzzy transition relations and fuzzy sets of
initial and terminal states which are represented by the following fuzzy matrices and vectors:
\[
\sigma^A=\begin{bmatrix} 1 & 1 & 1 \end{bmatrix},\quad
\delta_x^A=\begin{bmatrix}
 1  &  1  &  0  \\
 1  &  1  &  0  \\
 0  &  0  & \frac12
\end{bmatrix},\quad
\tau^A= \begin{bmatrix}
1 \\ 1 \\ 1
\end{bmatrix}, \qquad
\sigma^B=\begin{bmatrix} 1 & 1 \end{bmatrix},\quad
\delta_x^B=\begin{bmatrix}
 1  & 0  \\
 0  & \frac12
\end{bmatrix},\quad
\tau^B= \begin{bmatrix}
1 \\ 1
\end{bmatrix}.
\]
Computing the sequence $\{\varphi_k\}_{k\in \mathbb{N}}$ for forward bisimulations by the formula (\ref{eq:alg.seq})
($w=fb$) we obtain that
\[
\varphi_k=\begin{bmatrix}
 1  & \tfrac1{2^{k-1}} \\
 1  & \frac1{2^{k-1}} \\
\frac1{2^{k-1}} &  1
\end{bmatrix}, \ \ \text{for each $k\in \mathbb{N}$}, \qquad
\varphi = \bigwedge_{k\in\mathbb{N}}\varphi_k =
\begin{bmatrix}
 1  &  0 \\
 1  &  0 \\
 0  &  1
\end{bmatrix}.
\]
According to Theorem \ref{th:inf}, $\varphi $ is the greatest forward bisimulation between fuzzy automata $\cal A$ and $\cal B$.
\end{example}

The last example shows that the finiteness of the subalgebra $\langle \im (\psi^w)\cup \bigcup_{x\in X}(\,\im (\delta^A_x)\cup \im(\delta^B_x)\,)\rangle $ of $\cal L$, which appears as an assumption in Theorem \ref{th:alg}, is sufficient for the finiteness of the sequence defined~by~(\ref{eq:alg.seq}), but it is not necessary.

\begin{example}\rm
Let us change $\sigma^A$, $\sigma^B$, $\tau^A$ and $\tau^B$ in the previous example to
\[
\sigma^A=\begin{bmatrix} 1 & 1 & 0 \end{bmatrix},\qquad \sigma^B=\begin{bmatrix} 1 & 0 \end{bmatrix}, \qquad
\tau^A= \begin{bmatrix} 1 \\ 1 \\ 0 \end{bmatrix},\qquad \tau^B= \begin{bmatrix} 1 \\ 0 \end{bmatrix}.
\]
Computing the fuzzy relations $\varphi_k$, $k\in \mathbb{N}$, using the formula (\ref{eq:alg.seq}), we obtain that
\[
\varphi_1=\varphi_2 = \begin{bmatrix}
 1  &  0 \\
 1  &  0 \\
 0  &  1
\end{bmatrix},
\]
and it is the greatest greatest forward bisimulation between fuzzy automata $\cal A$ and $\cal B$.

On the other hand, we have that the subalgebra $\langle \im (\psi^w)\cup \im (\delta^A_x)\cup \im(\delta^B_x)\,\rangle $ of $\cal L$ is not finite, since it contains $\frac1{2^k}$, for every $k\in\mathbb{N}$.

\end{example}

\section{Concluding remarks}

Simulations and bisimulations between fuzzy automata have been introduced and studied recently~in~\cite{CIDB.11}. They have been defined as solutions of particular systems of fuzzy relation inequalities.~Using the method for computing the greatest solutiuons of similar systems of fuzzy relation inequalities developed in \cite{ICB.10}, in the present paper we provide efficient algorithms for deciding whether there is a simulation or bisimulation of a given type between given fuzzy automata, and for computing the greatest simulation/bisimulation of this type, if it exists.

Related systems of fuzzy relation inequalities will be discussed in the general context in our further research.

\end{document}